\documentclass[sigconf,authorversion]{acmart}

\AtBeginDocument{%
  \providecommand\BibTeX{{%
    \normalfont B\kern-0.5em{\scshape i\kern-0.25em b}\kern-0.8em\TeX}}}

\copyrightyear{2023} 
\acmYear{2023} 
\setcopyright{acmlicensed}\acmConference[IUI '23]{28th International Conference on Intelligent User Interfaces}{March 27--31, 2023}{Sydney, NSW, Australia}
\acmBooktitle{28th International Conference on Intelligent User Interfaces (IUI '23), March 27--31, 2023, Sydney, NSW, Australia}
\acmPrice{15.00}
\acmDOI{10.1145/3581641.3584052}
\acmISBN{979-8-4007-0106-1/23/03}

\usepackage{subcaption}
\usepackage{enumitem, hyperref}
\usepackage{siunitx}

\begin{document}


\title[The Effect of AI Delegation on Human Task Performance and Task Satisfaction]{Human-AI Collaboration: The Effect of AI Delegation on Human Task Performance and Task Satisfaction}

\author{Patrick Hemmer}
\affiliation{%
  \institution{Karlsruhe Institute of Technology}
  \city{Karlsruhe}
  \country{Germany}
  }
\email{patrick.hemmer@kit.edu}

\author{Monika Westphal}

\affiliation{%
 \institution{Ben-Gurion University of the Negev}
 \city{Be'er Sheva}
 \country{Israel}
 }
\email{monika.westphal@post.bgu.ac.il}

\author{Max Schemmer}
\affiliation{%
  \institution{Karlsruhe Institute of Technology}
  \city{Karlsruhe}
  \country{Germany}
  }
\email{max.schemmer@kit.edu}

 \author{Sebastian Vetter}
\affiliation{%
  \institution{Karlsruhe Institute of Technology}
  \city{Karlsruhe}
  \country{Germany}
}
\email{sebastian.vetter@alumni.kit.edu}

\author{Michael Vössing}
\affiliation{%
  \institution{Karlsruhe Institute of Technology}
  \city{Karlsruhe}
  \country{Germany}
  }
\email{michael.voessing@kit.edu}

\author{Gerhard Satzger}
\affiliation{%
  \institution{Karlsruhe Institute of Technology}
  \city{Karlsruhe}
  \country{Germany}
  }
\email{gerhard.satzger@kit.edu}


\renewcommand{\shortauthors}{Hemmer et al.}

\begin{abstract}
Recent work has proposed artificial intelligence (AI) models that can learn to decide whether to make a prediction for an instance of a task or to delegate it to a human by considering both parties' capabilities. In simulations with synthetically generated or context-independent human predictions, delegation can help improve the performance of human-AI teams---compared to humans or the AI model completing the task alone. However, so far, it remains unclear how humans perform and how they perceive the task when they are aware that an AI model delegated task instances to them. In an experimental study with 196 participants, we show that task performance and task satisfaction improve through AI delegation, regardless of whether humans are aware of the delegation. Additionally, we identify humans' increased levels of self-efficacy as the underlying mechanism for these improvements in performance and satisfaction. Our findings provide initial evidence that allowing AI models to take over more management responsibilities can be an effective form of human-AI collaboration in workplaces.
\end{abstract}



\maketitle

\section{Introduction}

Over the last few years, the capabilities of artificial intelligence (AI) have undergone considerable technical advances. Nowadays, the performance of AI models is similar to, and in certain application areas even exceeds the performance of human experts \cite{brown2019superhuman,he2015delving,silver2018general}. For example, in the medical domain, AI models can detect certain diseases as accurately as radiologists \cite{esteva2017dermatologist,gulshan2019performance,irvin2019chexpert}. Yet, despite these impressive advances, human predictions often remain more accurate for certain cases \cite{geirhos2021partial,wilder2020learning}. On the one hand, this may be due to limited model capacity, limited training data, or outliers unknown to the AI model. On the other hand, humans might have access to side information that is not readily available to the AI model, enabling them to make more accurate decisions for particular cases \cite{hemmer2022effect}. 
\begin{figure*}%
    \centering
    \subfloat[\centering Training of the AI model.]{{\includegraphics[width=0.48\textwidth]{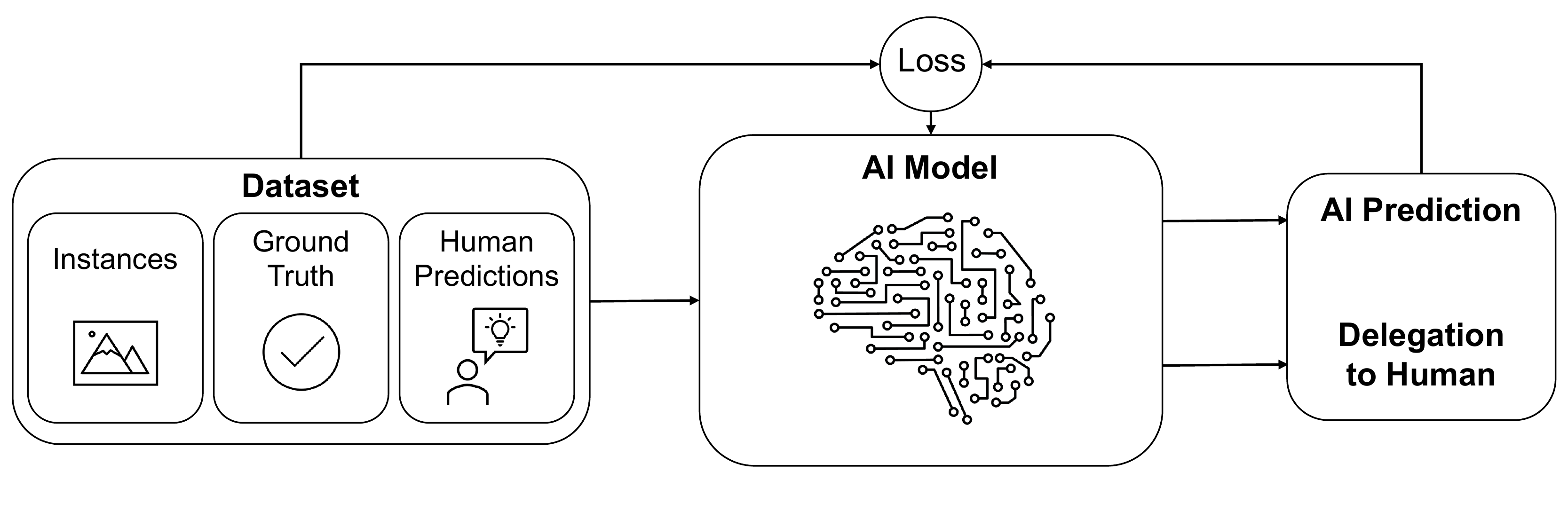} }}%
    \subfloat[\centering Deployment of the AI model.]{{\includegraphics[width=0.48\textwidth]{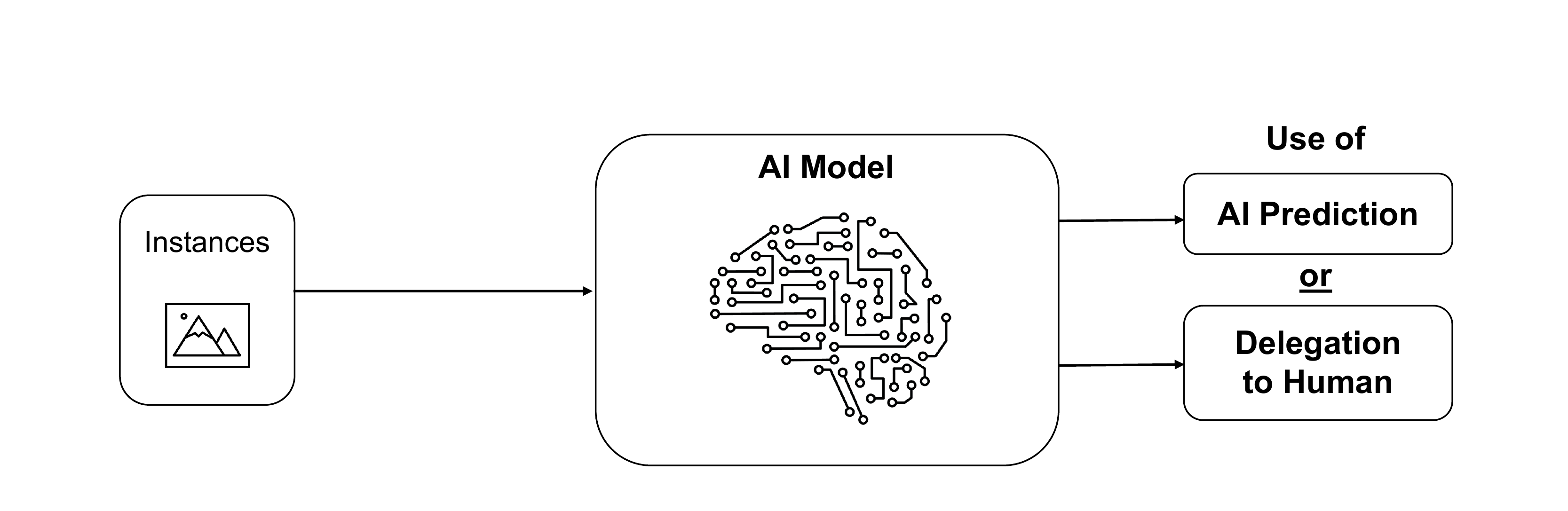} }}%
    \caption{A schematic overview of the AI model. During training (a), the AI model learns to make a prediction for a task from the available ground truth labels. Additionally, human predictions allow the AI model to learn the capabilities of humans simultaneously. After deployment (b), the AI model decides to make a prediction or to delegate an instance to the human, depending on whether the AI model or the human is expected to make a correct prediction with higher probability.}%
    \label{fig:example_ai_delegation}%
    \Description{A schematic overview of the AI model. During training (a), the AI model learns to make a prediction for a task from the available ground truth labels. Additionally, human predictions allow the AI model to learn the capabilities of humans simultaneously. After deployment, (b) the AI model decides to make a prediction or to delegate an instance to the human, depending on whether the AI model or the human is expected to make a correct prediction with higher probability.}
\end{figure*}
These potentially complementary capabilities motivated researchers to investigate how the abilities of humans and AI models can be combined to further improve overall decision-making performance \cite{Bansal2021,hemmer2schemmer021}. 

One noteworthy form of collaboration is the delegation of instances to a human by the AI model (i.e., AI delegation) \cite{leitao2022human}. Figure \ref{fig:example_ai_delegation} provides a schematic overview of AI delegation. This approach is particularly beneficial in application areas where tasks can be completed independently by both humans and AI models (e.g., crowdworking tasks like image recognition or content moderation) \cite{Desmond_2022,Lai2022}. AI delegation could also be used in high-stakes decision-making domains (e.g., medicine) to reduce the workload of medical experts. For instance, in the context of cancer screening, the AI model can be used to identify simple cases so that the medical experts can focus on the complex cases delegated to them \cite{bilal2022ai}. Several recent works propose approaches that enable the AI model to delegate a subset of instances to a human while taking both its own and the human's capabilities into consideration \cite{ijcai_multiple_experts,keswani2021towards,mozannar2020consistent,okati2021differentiable,Raghu2019,wilder2020learning}. One way to achieve this is to estimate both the AI model and the human prediction confidence on an instance basis and to delegate each instance to the team member with the higher estimated prediction confidence \cite{Raghu2019}. Generally, these works assume that the behavior and perceptions of humans, and thus their decision-making performance, remain unchanged whether or not an AI model delegates instances of a task. Previous research has demonstrated the potential of these approaches in experiments with either synthetically generated human predictions or with predictions that were collected in annotation settings without any AI involvement. However, human behavior might deviate when teaming up with an AI model. For example, humans' attitudes towards AI, their experience with algorithms, or exposure to an AI model that determines their task are factors that can influence their decision-making performance \cite{Bondi2022,dietvorst2015algorithm,Rastogi2022}. Thus, it remains an open question of how and why humans' performance is affected when the task is determined by an AI model. Following this, the first goal of this study is to investigate the effect of AI delegation on human task performance and to explore what drives this effect. 

Besides performance, the collaboration with an AI model that determines the task for humans might also have an effect on their perception of the work and the nature of the task. Human task satisfaction plays an increasingly important role in today's workplaces. In particular, humans' satisfaction with their work determines key organizational outcomes, e.g., commitment to or productivity of an organization, and is thus decisive for its long-term success \cite{Gerdenitsch2017,sadeghian2022artificial}. By delegating instances to the human, the AI model determines the nature of the task the human has to conduct, potentially affecting their satisfaction. Therefore, we investigate the effect of delegation by an AI model on human task satisfaction besides task performance, as well as what drives this effect (i.e., the underlying mechanism). We hypothesize improvements in both task performance and task satisfaction following AI delegation. Further, we expect increases in self-efficacy, i.e., a person's belief in one's own ability to complete a task successfully \cite{Bandura1977} to explain these positive effects. To summarize, in this work, we pose the following three research questions:
\begin{itemize}
    \item[\textbf{RQ1:}] \textit{How does AI delegation affect task performance compared to a human and an AI working alone?}
    \item[\textbf{RQ2:}] \textit{How does AI delegation affect task satisfaction compared to a human working alone?}
    \item[\textbf{RQ3:}] \textit{What explains the effect of AI delegation on task performance and task satisfaction?}
\end{itemize}

To answer these research questions, we conduct a randomized experiment with 196 participants recruited online via Prolific. Participants are asked to complete an image classification task based on a modified subset of the ImageNet data set \cite{ILSVRC15,Steyvers2022}. We select this task because it does not require any task-specific training to achieve similar performance compared to modern AI models \cite{Steyvers2022}. We employ an AI model that learns to classify images and simultaneously estimates the instance-specific human classification confidence that is compared with the confidence of the classifier. Instances are delegated if the estimated human confidence is higher than the confidence of the classifier \cite{Raghu2019}. The experiment includes two ``delegation'' treatment groups that receive a randomly drawn subset of the images in the test set that the trained AI model had selected for delegation to humans. In one of the groups, humans are aware of the AI delegation taking place, while in the other, humans classify the same images without knowing about the AI delegation. We investigate the effect of AI delegation on task performance and task satisfaction in these two delegation groups compared to a control group (i.e., ``human-alone''), where humans classify a subset of images randomly selected from the test set. In addition, we compare the performance of these groups with the performance of the AI model if it had conducted the task alone. We find that humans' performance increases significantly for the instances delegated by the AI model, which results in an overall team performance exceeding the performance of both humans and the AI conducting the task independently. Additionally, we find that humans' task satisfaction increases significantly when they work on the delegated set of instances. Both effects can be explained by an increase in humans' self-efficacy. Interestingly, we find no differences in human task performance and task satisfaction, regardless of whether humans are informed about the AI delegation. Thus, we can conclude that the modified nature of the task drives the observed positive effects of delegation through the AI model. All these findings show the potential of AI delegation as an appropriate form of collaboration between humans and AI.

To summarize, our contributions are as follows: (1) We propose a behavioral model to analyze the effect of AI delegation on human task performance and task satisfaction in human-AI collaboration. (2) We validate our model in a randomized experiment with human participants and show that their performance and task satisfaction are improved through the delegation of instances by the AI model. Moreover, we show that the overall team performance surpasses the individual performance of both team members working alone. (3) We identify self-efficacy as an underlying mechanism to the effect of AI delegation on task performance and on task satisfaction.

\section{Related work}

The collaboration between humans and AI models can be instantiated in different ways. One of the most common collaboration forms between humans and AI is AI-assisted decision-making---a setting in which an AI model provides recommendations to support the human. The human is in the role of making the final decision and can, therefore, either accept or override the recommendation \cite{Schemmer2022_meta,wang2021explanations}. Establishing an appropriate level of reliance on the AI model becomes one of the central challenges \cite{schemmer2022should}. Thus, the AI model often provides the confidence level of the decision \cite{nguyen2021effectiveness,zhang_efect_2020} or an additional explanation for its decision \cite{adadi_peeking_2018,lai_human_2019}. Several works have evaluated whether different types of explanations can support humans' understanding of the AI model so that they identify the right cases to rely on the recommendations \cite{alufaisan_does_2020,bucinca_proxy_2020,carton_feature-based_2020,van_der_waa_evaluating_2021}. Explanations can lead people to rely too much on the decision of the AI model, particularly when its suggestion is incorrect \cite{Bansal2021}. This over-reliance also depends on the humans' level of task-specific expertise. For example, people with higher task expertise become more confident in overruling the recommendation of the AI model \cite{feng2019can}. Wrong predictions, recognized as such by humans, can lead to a loss of trust in the system \cite{dietvorst2015algorithm,nourani2020role}. Recent research investigates other factors that might play an essential role in AI-assisted decision-making, e.g., whether humans' performance benefits from receiving tutorials about the functionality of the AI model \cite{lai_why_2020} or whether specific design elements can foster people's engagement with AI explanations \cite{bucinca_trust_2021}.

Besides AI-assisted decision-making, a different type of human-AI collaboration has attracted increasing interest in research over the past few years---delegation initiated by the AI model (i.e., AI delegation) \cite{leitao2022human}. The AI model learns to decide whether to make a prediction itself for a given task instance or to delegate it to a human. In application domains with a high number of individual decisions delegating instances of a task can reduce human effort and improve overall performance. Instances are distributed to the team member who is most likely to make the correct decision. Typically, these approaches take not only their own but also the capabilities of the humans into consideration \cite{ijcai_multiple_experts,keswani2021towards,mozannar2020consistent,okati2021differentiable,Raghu2019,wilder2020learning}. The AI model learns the strengths and weaknesses of the human team member from human predictions used during model training in addition to the ground truth labels. Such individual human predictions are noisy compared to the ground truth labels. The latter are typically determined by experts or multiple individual human predictions to ensure high label quality \cite{Kerrigan2021}. Different algorithms have been proposed that can either complement the capabilities of a single \cite{mozannar2020consistent,okati2021differentiable,Raghu2019,wilder2020learning} or multiple \cite{ijcai_multiple_experts,keswani2021towards} humans who are part of the human-AI team. So far, these approaches have solely been evaluated with synthetically generated or context-independent human predictions that were collected in annotation settings without any AI involvement. However, human predictions might deviate when they are aware of the AI delegation taking place, e.g., due to their attitude or prior experience with algorithms or due to being attributed with particular competence by the AI model that takes on the role of a manager \cite{Bondi2022,dietvorst2015algorithm}. Therefore, it remains an open question whether humans' individual performance and the overall team performance in real-world settings would benefit from delegation algorithms that consider both parties' capabilities. Furthermore, research has so far neglected the possible effect on humans’ perceptions of being managed by the AI model, e.g., expressed through task satisfaction. However, task satisfaction plays a central role in workplaces where people are increasingly exposed to working with AI models, especially when they decide on the task a human has to complete. Research lacks an understanding of the underlying mechanisms of the effect of AI delegation on task performance and task satisfaction in human-AI collaboration. Only \citet{Bondi2022} and \citet{fuegener2022cognitive} investigated AI delegation in behavioral experiments. However, these studies differ in two ways from the current study: First, the algorithms used for delegation do not learn the capabilities of the humans. Second, they do not investigate humans' perceptions when the AI model delegates task instances, nor do they aim to understand the underlying mechanisms driving the effects on task performance and task satisfaction.

\section{Theory Development and Hypotheses}
\label{sec:Theoretical development}

So far, AI models that learn to decide whether to make a prediction themselves or to delegate an instance to a human have been evaluated with synthetically generated or context-independent human predictions \cite{ijcai_multiple_experts,keswani2021towards,mozannar2020consistent,okati2021differentiable,Raghu2019,wilder2020learning}. However, humans' behavior may deviate when they are aware that they are part of a human-AI team \cite{Bondi2022,dietvorst2015algorithm}. It remains unclear how such a team setting would affect human performance and other individual task outcomes (e.g., task satisfaction) in real-world settings. Furthermore, it is not yet known, why AI delegation may affect individual task outcomes. In this study, we examine how AI delegation affects task performance and task satisfaction, considering self-efficacy as a possible underlying mechanism. We draw upon experimental studies in organizational behavior literature on the effect of supervisor-to-employee delegation, and its relation to task performance, satisfaction, and self-efficacy \cite{Locke1984,schriesheim1998delegation,shore1989job}. Based on this literature, we develop four hypotheses that are subsequently tested in an experimental study. 

Research in organizational behavior suggests that delegation from a supervisor to an employee can serve multiple purposes. For example, supervisors delegate due to a lack of time, missing competencies, or to empower employees for their personal development \cite{bass1990bass,schriesheim1998delegation}. Several works identified a positive relationship between supervisor delegation and employee performance \cite{al2015impact,Leana1986,leana1987power,xiong2007delegation}. When aligned with the employees' competencies, delegation results in more empowered, motivated, and higher-performing employees \cite{ugoani2020}. We transfer these insights to the modern context of human-AI collaboration. We propose a positive effect of AI delegation in human-AI collaboration, given the AI model learns the strengths and weaknesses of the human collaborator:

\begin{itemize}
    \item[\textbf{H1:}] \textit{AI delegation improves human task performance compared to an AI and a human working alone.}
\end{itemize}

Organizational behavior research has investigated employee job satisfaction as another important factor besides performance; precisely because it determines key organizational outcomes such as employee organizational commitment and turnover \cite{shore1989job}. Several works identified delegation as positively related to employees' job satisfaction \cite{farrow1980comparison,ugoani2020,xiong2007delegation}. For example, \citet{schriesheim1998delegation} found that perceived delegation by employees improved their intrinsic and extrinsic job satisfaction. We take these insights to the modern context of AI delegation in human-AI collaboration and propose the following effect, given the AI model delegates instances to the humans that align with their competencies: 

\begin{itemize}
    \item[\textbf{H2:}] \textit{AI delegation improves human task satisfaction compared to a human working alone.}
\end{itemize}

Besides examining the direct effect of AI delegation on task performance and task satisfaction, we aim to understand why these proposed effects occur. We investigate self-efficacy as a potential underlying mechanism. Self-efficacy refers to the confidence in one's own ability to complete a task successfully \cite{Bandura1977}. Again drawing upon experimental studies in organizational research, we find that delegation from a supervisor to an employee enhances psychological empowerment \cite{zhang2017leaders}. In other words, delegation makes employees feel that their job is meaningful and that they are responsible for work outcomes. We are not aware of any study showing that delegation increases self-efficacy. Still, many studies are pointing to the role of (increased) self-efficacy in improving organizational performance-related outcomes. For example, self-efficacy influences learning and the level of effort put into work \cite{lunenburg2011self}. Further, self-efficacy predicts several work-related performance outcomes \cite{stajkovic1998self, Locke1984}. In a learning environment, self-efficacy correlates with increased task performance \cite{Afzal2019}. Besides performance, self-efficacy improves job satisfaction through higher meaningfulness \cite{baumeister2002,gecas1982}. Organizations should select potential employees based on their self-efficacy levels; employees with high self-efficacy levels are more motivated and more likely to yield desired outcomes for the organization \cite{lunenburg2011self}. Self-efficacy has important implications for organizational behavior and human resource management \cite{gist1987self}. 

In the current study, we are interested in understanding self-efficacy in a modern work context---human-AI collaboration, where an AI model delegates task instances. The hope is to yield higher levels of task performance and task satisfaction. Based on the findings of experimental studies in organizational behavior research mentioned above, we propose an additional set of hypotheses:

\begin{itemize}
    \item[\textbf{H3:}] \textit{Self-efficacy mediates the effect of AI delegation on human task performance. In particular, AI delegation increases self-efficacy, and this increased self-efficacy improves task performance compared to a human working alone.}
\end{itemize}

\begin{itemize}
    \item[\textbf{H4:}] \textit{Self-efficacy mediates the effect of AI delegation on human task satisfaction. In particular, AI delegation increases self-efficacy, and this increased self-efficacy improves task satisfaction compared to a human working alone.}
\end{itemize}

Figure \ref{Fig:research_model} provides an overview of our research model and proposed effects:

\begin{figure}[htb]
  \centering
  \includegraphics[width=\linewidth]{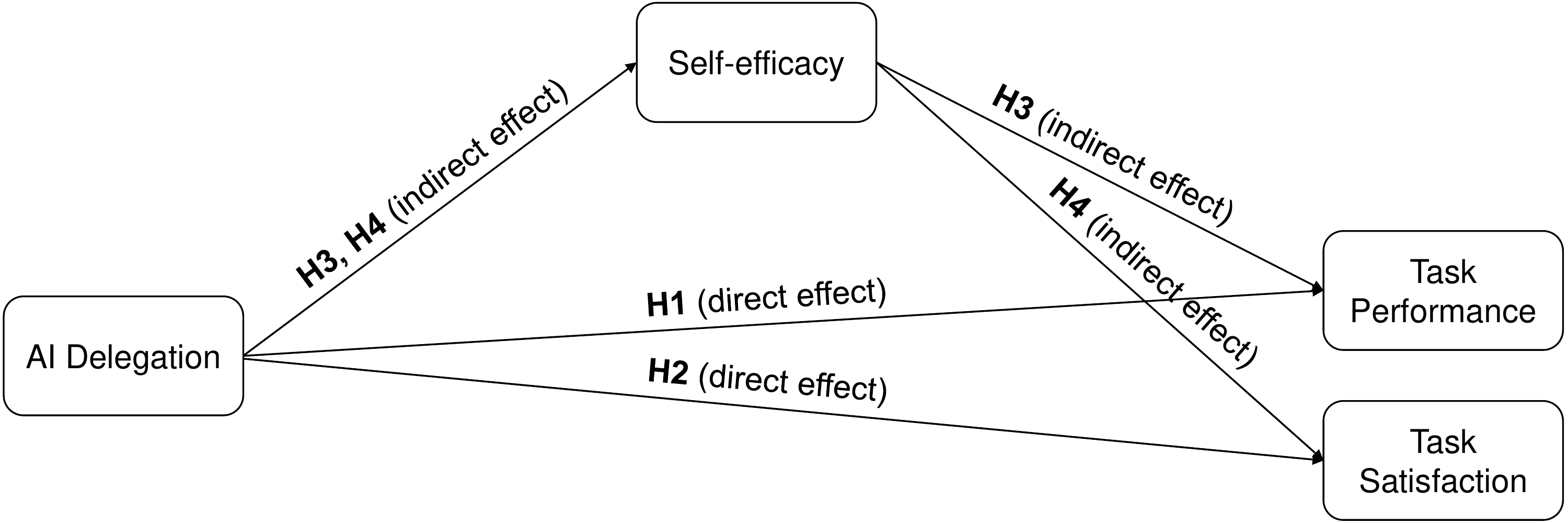}
  \caption{Research model: Proposed effects of AI delegation on human task performance and satisfaction.}
  \label{Fig:research_model}
  \Description{Research model: Proposed effects of AI delegation on human task performance and satisfaction.}
\end{figure}

As stated in the hypotheses, we compare the task outcomes for humans working on instances delegated by an AI (i.e., AI delegation) to two control groups where no delegation takes place (i.e., humans working alone; AI working alone). 

In addition to the proposed effect of AI delegation on task performance and task satisfaction, we examine another exploratory research question: Does AI delegation affect task performance and task satisfaction differently when the delegation is \textit{not} communicated to the human? In other words, the human works (only) on the instances delegated by the AI model but is not informed about it. To test this exploratory research question, we include a second delegation group in the design, the ``hidden delegation'' group. Though the task experience might be similar to when working alone, we think the human might experience the task more positively because the delegation algorithm works well by delegating the right instances. The question is whether humans will experience the task even more positively than those in the delegation group who are informed about the delegation. In any case, we expect to see a positive effect of AI delegation on task performance and task satisfaction. Next, we outline how we tested our propositions in an experimental study.

\section{Methodology}

In this section, we first provide information about the data we used. Then, we describe the development of the AI model. Finally, we present the experiment that we conducted to test the hypotheses. 

\subsection{Data}\label{sec:data}

We used the image data set provided by \citet{Steyvers2022} for our study. The data set is a subset of the ImageNet Large Scale Visual Recognition Challenge (ILSVRC) 2012 database \cite{ILSVRC15}. It consists of 1,200 images equally balanced over 16 classes, e.g., airplane, bear, or boat. Additionally, phase noise distortion was applied to the images at each spatial frequency, uniformly distributed in the interval \([\omega,\omega]\) with \(\omega=110\) to increase the difficulty of the classification task both for humans and the AI model. Despite the increased difficulty level, both humans and the AI model can achieve a similar performance level on the task. We refer to \citet{Steyvers2022} for additional details. We chose this data set as a test bed for our proposed behavioral model for multiple reasons: First, it includes a generic, non-specialized task that can be conducted by non-specialized participants. Hereby, we aim to ensure a certain degree of generalizability of the results. Second, in addition to the ground truth labels, the data set provides multiple human predictions for each image collected from 145 Amazon Mechanical Turk workers. Thus, it fulfills our requirement that the AI model can learn the humans' strengths and weaknesses. 

We prepared the data set by randomly selecting a human prediction for each image that is subsequently used together with the ground truth labels for the training of the AI model. We divided the data set into a training, validation, and test set, with 60\%, 20\%, and 20\% of the data, respectively. 

\begin{figure*}[h]
    \centering
    \subfloat[\centering Exemplary image of the classification task.]{{\includegraphics[width=0.46\textwidth]{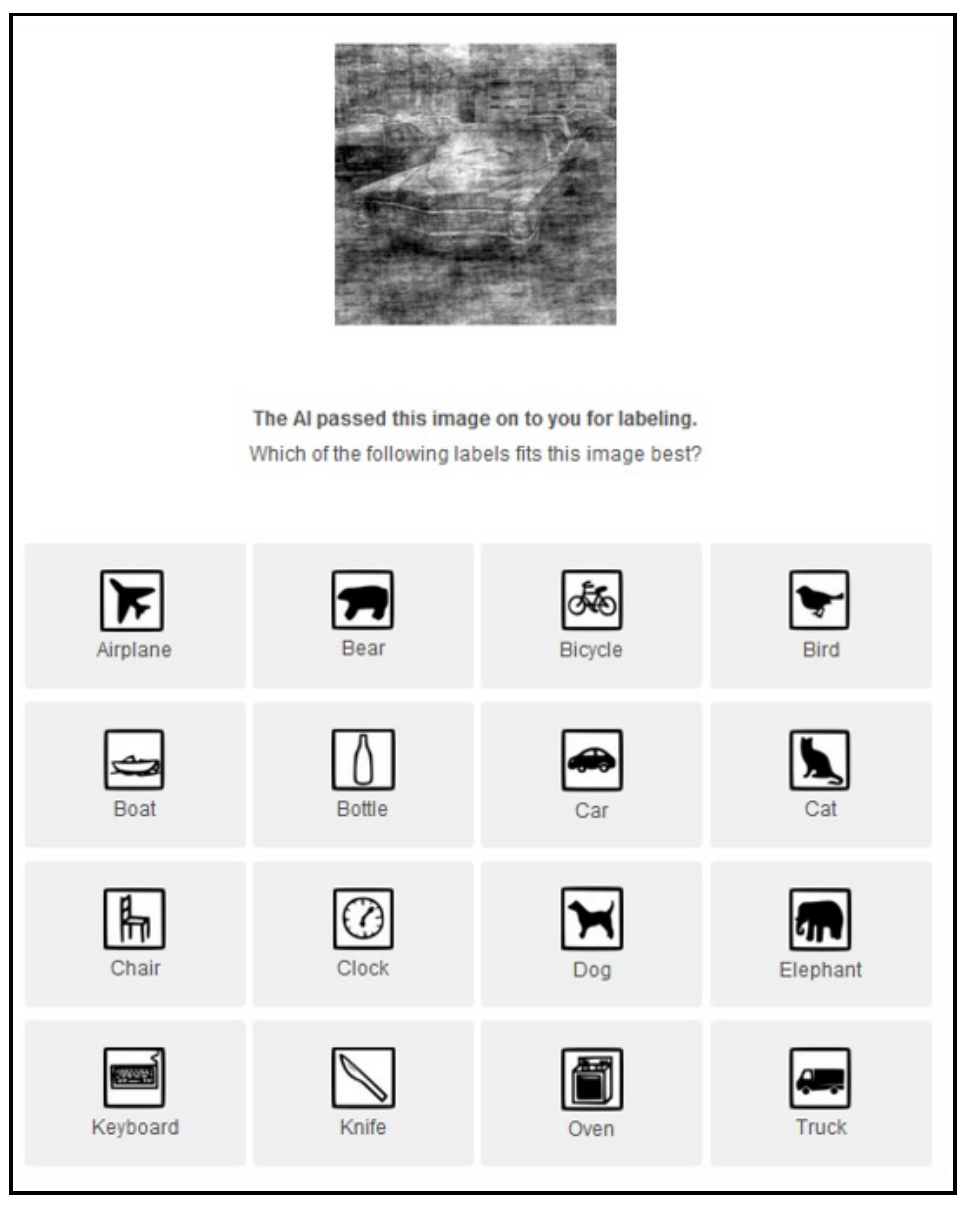}
    }\label{Fig:task_a}}%
    \subfloat[\centering Exemplary additional image classified by the AI model.]{{\includegraphics[width=0.46\textwidth]{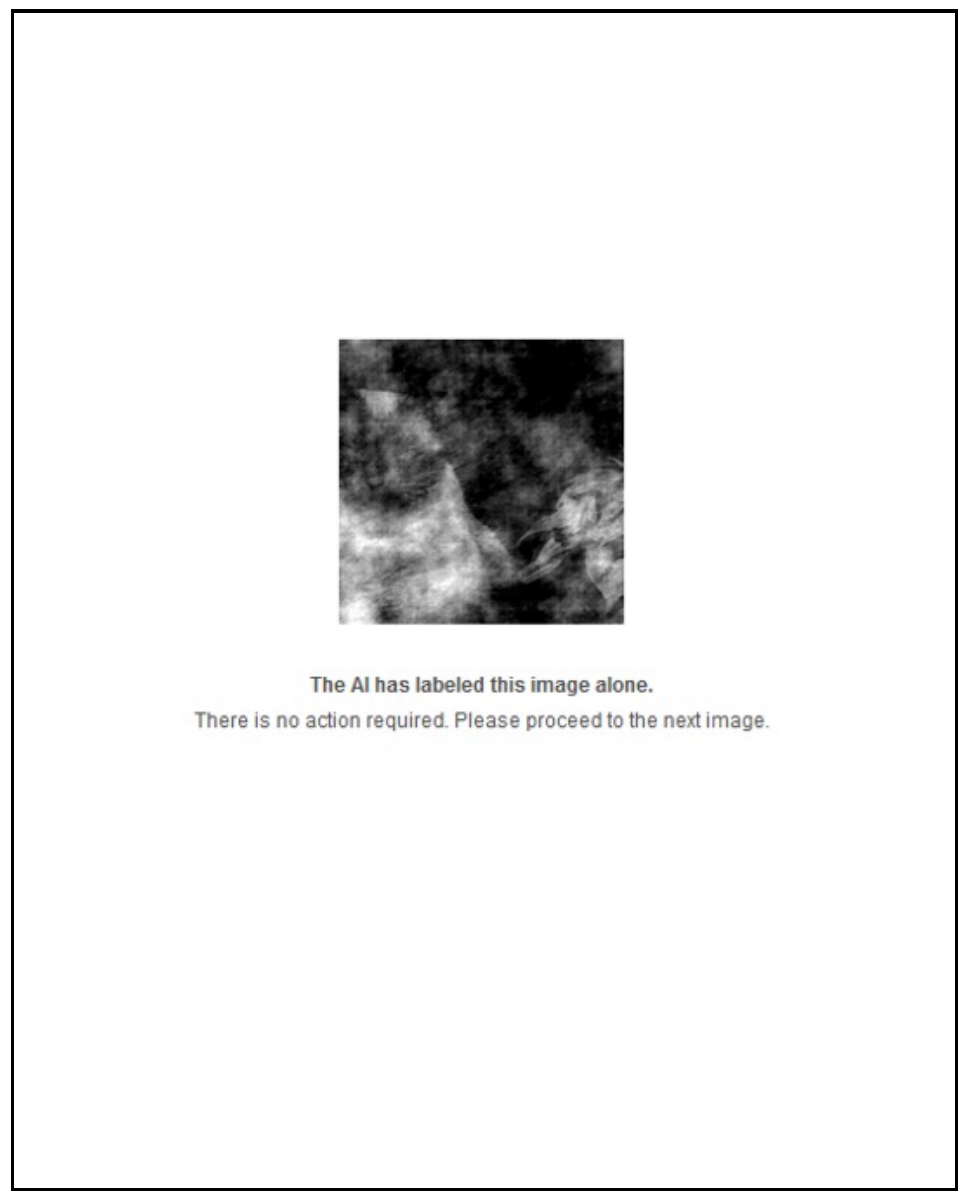}
    }\label{Fig:task_b}}%
    \caption{Interface of the image classification task, exemplified by the delegation condition. Participants were informed that the AI model delegated the respective image to them. Additionally, between the individual instances, participants saw images that the AI model had already classified and, thus, did not delegate to them.}%
    \label{Fig:task}%
    \Description{Interface of the image classification task, exemplified by the delegation condition. Participants were informed that the AI model delegated the respective image to them. Additionally, between the individual instances, participants saw images that the AI model had already classified and, thus, did not delegate to them.}
\end{figure*}

\subsection{Development of the AI Model}
\label{sec:development}

For the AI model, we implemented the approach proposed by \citet{Raghu2019}. It consists of two components: First, a classification model that learns the image classification task. Second, a human error model that learns to predict whether humans would classify an instance correctly based on the provided human predictions. The delegation decision is made based on the instance-level confidence of both components. If the human error model has higher confidence than the classification model, the instance is subsequently delegated to the human. Both the classification and human error models consist of a DenseNet-161 \cite{huang2017} pre-trained on ImageNet. We fine-tuned both models on the distorted images over 100 epochs using SGD as an optimizer with a learning rate of \(1 \cdot 10^{-4}\), weight decay of \(5 \cdot 10^{-4}\), a cosine annealing learning rate scheduler and a batch size of 16. Additionally, we applied early stopping on the validation loss.

\subsection{Experimental Design}
\label{sec:experimental design}

To test the effect of AI delegation on human performance and satisfaction, as well as self-efficacy (see Hypotheses 1--4), we conducted a web-based experiment. Next, we outline the experimental design: participants, study procedure, and evaluation measures.

\subsubsection{Participants}

We calculated the required sample size using G-Power \cite{faul2007g} and assumed a small effect (0.10). Accordingly, 176 participants were necessary to detect effects between three groups in a multiple linear regression (including three predictors, fixed model, R$^2$ deviation from zero), with a power of 0.95. As it is common for some participants to fail attention or manipulation checks or drop out of the study, we recruited a larger number of participants (roughly 15\% on top of the calculated number). Following this, we recruited 210 participants online via Prolific Academic. Participants received \$1.5 for their participation in the task that took approximately 10 minutes. We excluded 13 participants because they failed the attention or manipulation check, and an additional participant due to missing data. Hence, our final sample was 196 participants ($Mean=39.43$ years, $SD=13.13$; 58.67\% female).

\subsubsection{Study Procedure}

At the beginning of the study, participants were asked to perform an ``unrelated task'' that estimated their cognitive ability to handle visual cognitive tasks. We included this variable as a control in our analyses. Next, participants had to pass an attention check. Following that, they started the practice round of the main task: We asked them to classify three images, one after another, to familiarize themselves with the task. Participants had to choose from a four-by-four matrix including 16 icons of the objects, each representing a different class, with the name of the class displayed underneath the icon (e.g., dog, airplane, truck). They saw the three images in random order and the 16 icons of the objects in alphabetic order. We chose the images randomly based on the test set as outlined in Section \ref{sec:data}. After the practice round, participants proceeded to the main task. They were asked to classify another 20 images. We randomly assigned them to three experimental conditions. 

In the \textit{delegation} condition, we informed them that ``this time, the AI will decide for each image whether to label it alone or to pass it on to you for labeling''. The 20 images the participants had to label were randomly drawn from the subset of images in the test set that the trained AI model had selected for delegation to humans. Moreover, we included five additional images and communicated that the AI had already labeled this image and that they could proceed to the next image. We neither mentioned the accuracy of the AI nor revealed the ground truth itself. Figure \ref{Fig:task} displays the interface presented to the participants. In detail, Figure \ref{Fig:task_a} shows an exemplary instance that was delegated to the participants by the AI model. Figure \ref{Fig:task_b} depicts one of the five additional images notifying the participants that the AI model has already labeled it. 

We included a second delegation condition in the design---the \textit{hidden delegation} condition. Just as in the delegation condition, participants were asked to label 20 images randomly drawn from the subset of images in the test set that the trained AI model had selected for delegation to humans. But this time, we did not communicate the delegation to the participants; we just told them: ``Just as previously, you will decide on the label for each image''. 

The third condition---the \textit{human-alone} condition---represented the control condition. Participants received the same information as in the \textit{hidden delegation} condition about the task. However, the 20 images were randomly drawn from the entire test set. 

Once participants had classified the 20 images, they responded to several follow-up questions that measured self-efficacy and task satisfaction, as well as recorded demographics and included a manipulation check.

\subsubsection{Evaluation Measures}\label{sec:evaluation_measures}

We measured the following variables to evaluate the effect of AI delegation: 

\textbf{\textit{Task performance.}} As instances are equally distributed over the 16 classes, task performance was measured by the percentage of correctly classified images, i.e., classification accuracy (\(acc\)). To assess the human performance, we calculated this measure for all three experimental conditions. Regarding the AI model, we calculated its performance on the test set and on the set of instances delegated to the humans.

Besides individual task performance, we were interested in the combined human-AI team performance. Hence, we also calculated the performance of the AI model on the subset of the test set not delegated to the humans (\(acc_{AI,\,\neg delegated}\)). We determined the team performance for each of the \(N\) participants in the delegation group by weighting their individual performance (\(acc_{human, \, delegated}\)) by the ratio of delegated images in the test set \(X\). Then, we combined it with the performance of the AI model weighted by the ratio of not delegated images in the test set \(1-X\). Lastly, we calculated the average team performance of all participants in the group. We refer to Equation \ref{eq:ctp} for this procedure:
\begin{equation}
    acc_{team} = \frac{1}{N} \sum_{i=1}^{N} \bigl( acc_{human,\,delegated}^{(i)} \cdot X + acc_{AI,\,\neg delegated} \cdot (1-X) \label{eq:ctp} \bigl)
\end{equation}

\textit{\textbf{Task satisfaction.}} We measured task satisfaction using three items based on \citet{Hofmann1995} on a validated, 5-point Likert scale (1 - `not at all' to 5 - `totally'). The three items were: 'Overall, how satisfied are you with your performance on this task?', `Overall, how satisfied are you with how much you learned?', and `Overall, how much did you enjoy performing this task?'. Cronbach's Alpha was 0.73 (sufficient).

\textit{\textbf{Self-efficacy.}} We measured self-efficacy using three items adapted from \citet{spreitzer1995} on a validated, 7-point Likert scale (1 - `I totally disagree' to 7 - `I totally agree'). The three items were: `I am confident about my ability to do the task.', `I have mastered the skills necessary for the task.', and `I am self-assured about my capabilities to perform the task.'. Cronbach's Alpha was 0.89 (good).

\textbf{\textit{Control variables.}}
We assessed the \textit{cognitive ability} to handle complex visual tasks, based on four items by \citet{Jacobs2014}. Participants were asked how easy or difficult they perceive certain tasks, compared to others of the same age, and evaluated the following four statements on a validated, 7-point Likert scale (1 - `extremely difficult' to 7 - `extremely easy'): `interpret visually displayed information', `understand information presented in a visual format', `imagine what an object would look like from a different angle', and `mentally rotate three-dimensional images in my mind'. Cronbach's Alpha was 0.86 (good). Besides \textit{cognitive ability}, we recorded participants' \textit{task experience}, \textit{algorithm attitude}, \textit{algorithm use}, \textit{education}, \textit{age}, and \textit{gender}.

\textbf{\textit{Manipulation checks.}}
We included the following statement in the description of the main task to make sure that participants in the delegation condition were aware of the delegation taking place: `Please show us that you have read the task description above by choosing the right response': (a) `Next, I will label all images alone (just as in the practice round)', (b) `either the AI or I will label the image', or (c) `the AI will label all images'. Participants were included in the analysis if they chose (b). Additionally, to make sure participants paid attention to the condition they were assigned to, we asked them at the end of the study whether (a) they `labeled all images alone, just as in the practice round', or (b) `an AI passed some of the images on to them for labeling', or (c) they `don't remember'. Only participants who chose (a) and were indeed either in the hidden delegation or the human-alone condition and participants who chose (b) and were indeed in the delegation condition were included in the final sample.

\begin{figure*}[t]
    \centering
    \resizebox{\textwidth}{!}{%
    \subfloat[\centering Task Performance.]{{\includegraphics[width=0.42\textwidth]{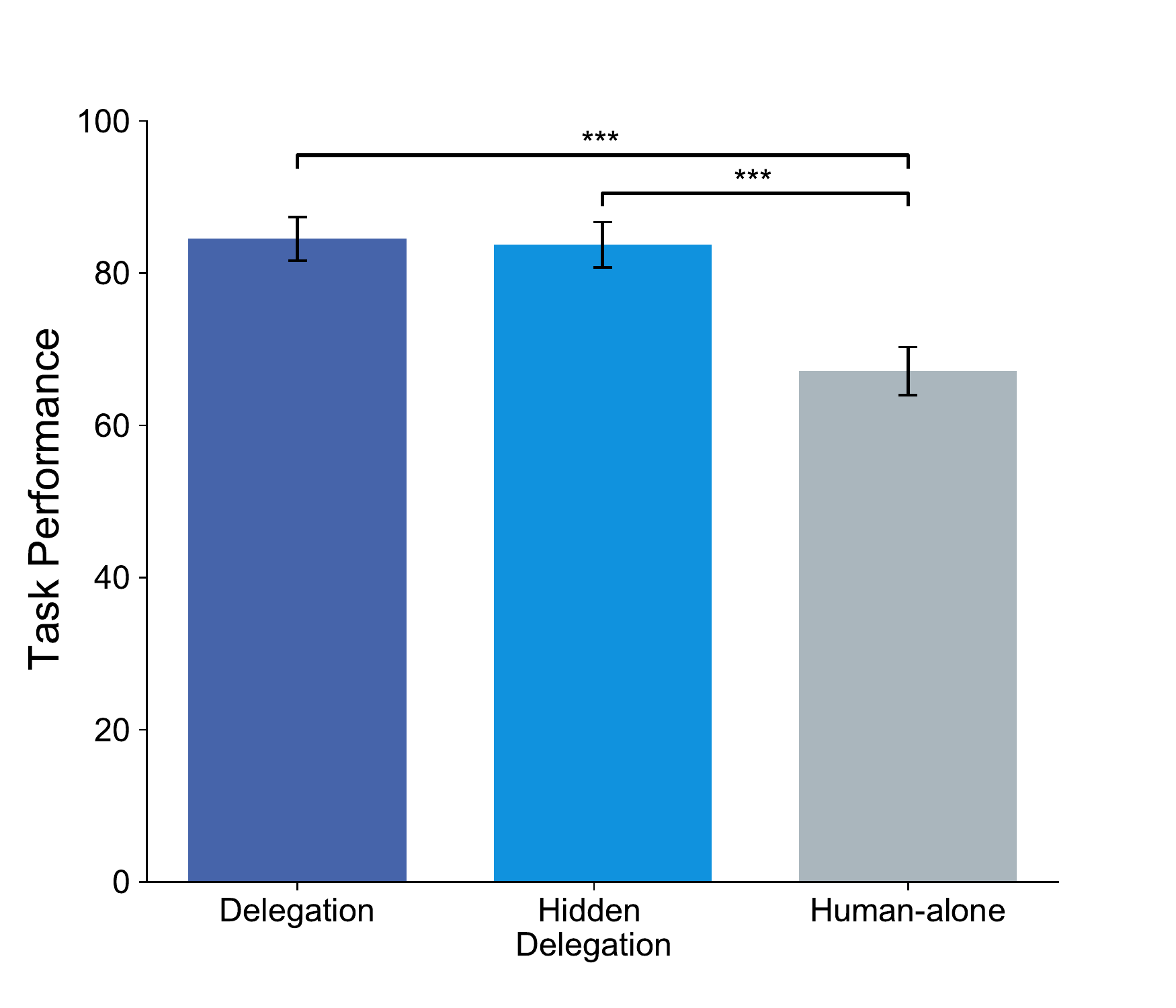}
    }\label{Fig:performance}}%
    \subfloat[\centering Task Satisfaction.]{{\includegraphics[width=0.42\textwidth]{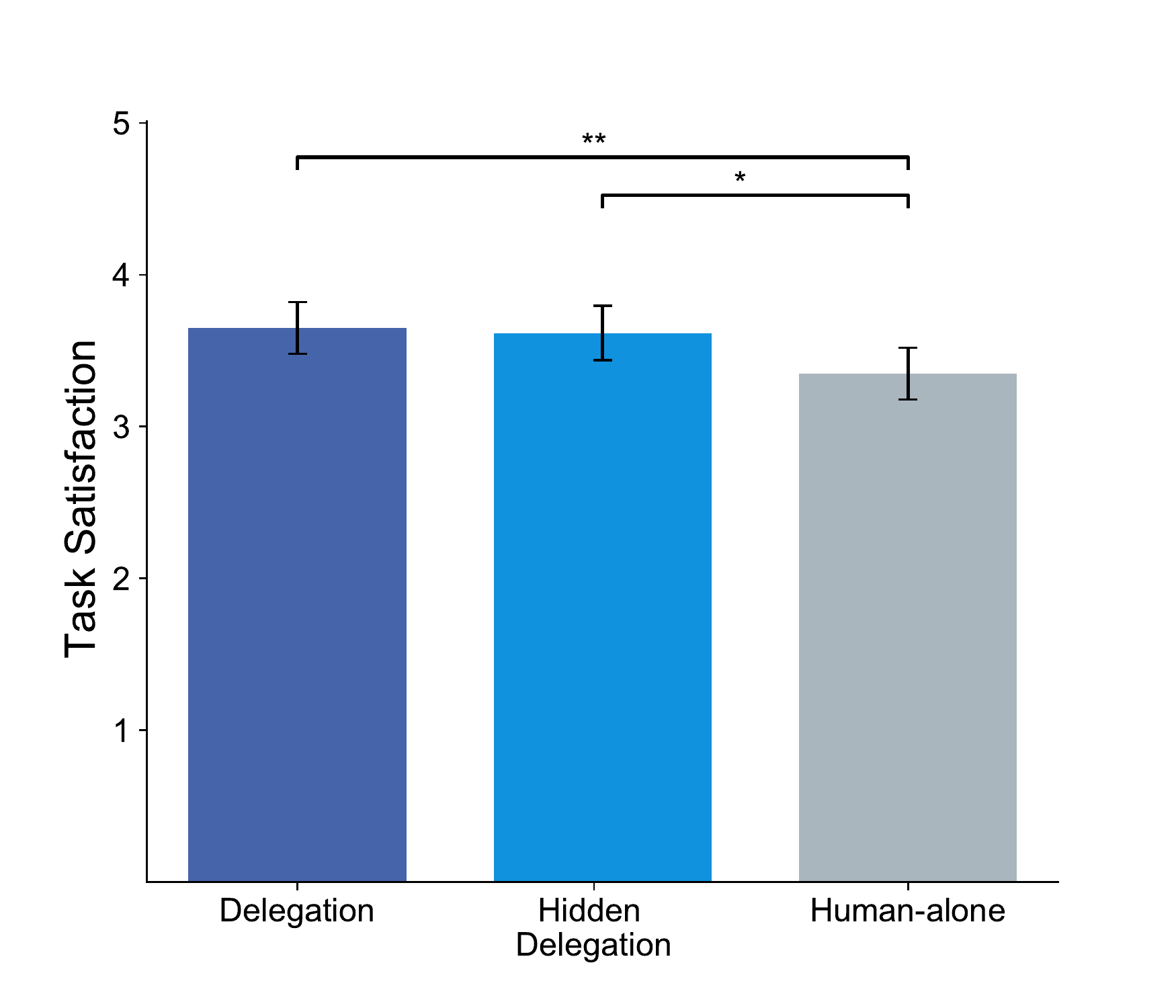}
    }\label{Fig:satisfaction}}%
    \subfloat[\centering Self-efficacy.]{{\includegraphics[width=0.42\textwidth]{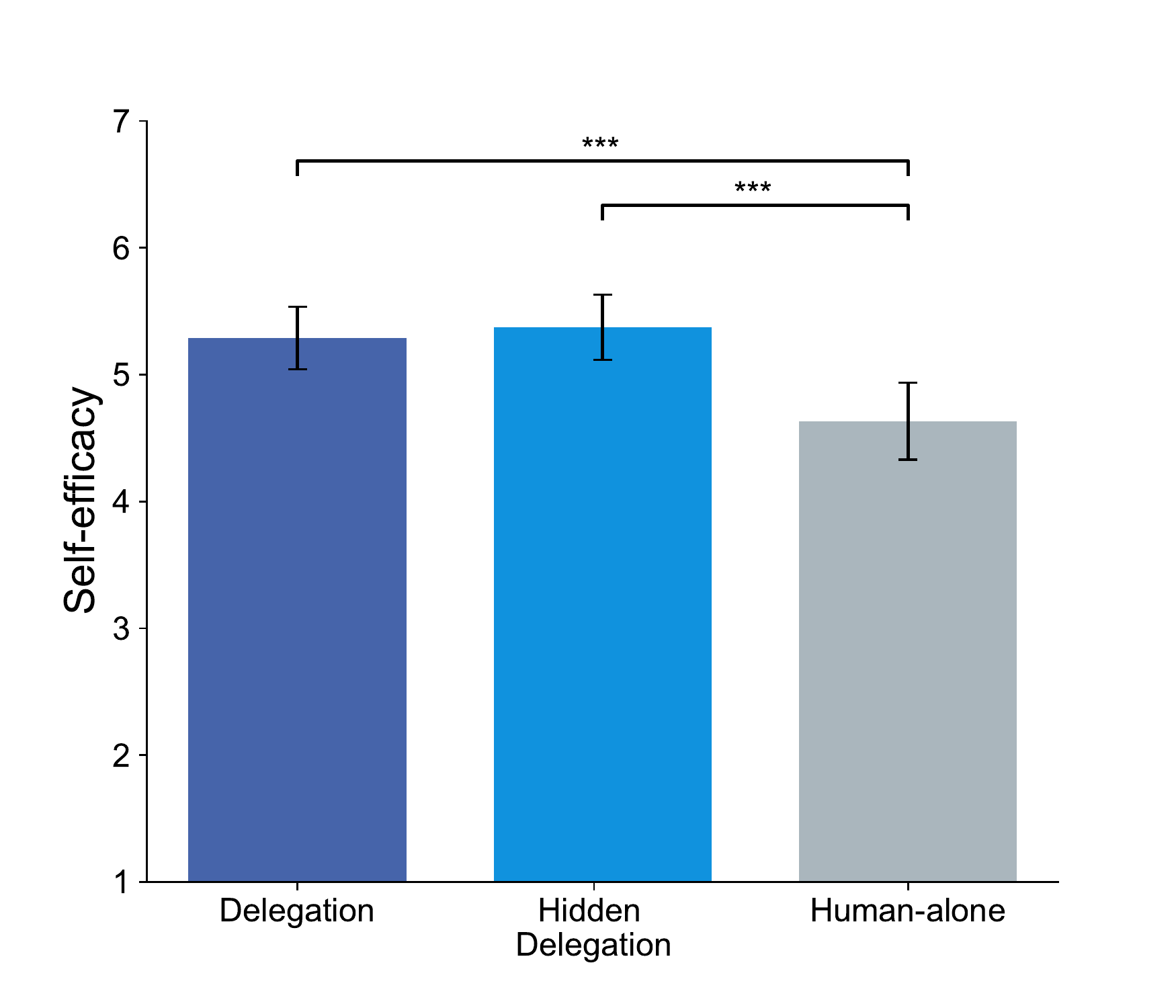}
    }\label{Fig:self_efficacy}}%
    }
    \caption{Task performance, task satisfaction, and self-efficacy of the participants, split by conditions. All bars include 95\% confidence intervals. Note: $^{***}$ $p<0.001$; $^{**}$ $p<0.01$; $^{*}$ $p<0.05$.}%
    \label{Fig:results_performance}%
    \Description{Task performance measured in accuracy, task satisfaction, and self-efficacy of the participants split by conditions. All bars include 95\% confidence intervals. Note: $^{***}$ $p<0.001$; $^{**}$ $p0<.01$; $^{*}$ $p<0.05$.}
\end{figure*}

\section{Results}
\label{sec:results}
\subsection{Statistical Specification} 
Our experiment examined four hypotheses that we developed in Section \ref{sec:Theoretical development}. The first set of hypotheses regarded the effect of AI delegation on task performance (H1) and task satisfaction (H2). The second set of hypotheses regarded the mediating role of self-efficacy in the effect of AI delegation on task performance (H3) and task satisfaction (H4). To test Hypotheses 1 and 2 (direct effect of AI delegation), we ran a univariate regression analysis that predicted task performance, and another one that predicted task satisfaction. To test Hypotheses 3 and 4 (indirect effect of AI delegation), we ran a mediation analysis for each of the two outcomes, using PROCESS \cite[model no.\ 4,][]{Hayes2017IntroductionApproach}, and including the mediation indices. 

In all analyses, we included all control variables, i.e., participants' task experience, algorithm attitude, algorithm use, cognitive ability, education, age, and gender.

\subsection{Effect of AI Delegation on Task Performance}

The overall regression model---testing for the direct effect of AI delegation on task performance---is significant, $F(9,186)=11.817, R^2=0.364, p<0.001$. As hypothesized, participants in both the delegation group and hidden delegation group yield higher levels of task performance ($Mean=84.51\%, SD=11.24, p<0.001$ and $Mean=83.73\%, SD=12.29, p<0.001$, respectively), compared to humans working alone ($Mean=67.13\%, SD=13.11$). A Tukey post hoc test reveals no significant difference in task performance between the two delegation groups ($p=0.932$). Figure \ref{Fig:performance} displays the performance results of all three groups. We observe that participants with a more positive attitude towards algorithms, and those who are younger, perform better ($p=0.009$ and $p=0.008$, respectively). For an overview of the regression results, see Figure \ref{Fig:results_mediation} and Table \ref{tab:results_regression} (Columns: `Model I---Direct effect of AI delegation', `Task performance').

Additionally, we compare the task performance of the delegation group with the performance of the AI model if it had classified the images presented to the delegation group alone. The accuracy of the delegation group is significantly higher ($p<0.001$, one-sample, one-tailed Wilcoxon signed rank test) than the accuracy of the AI model on the delegated set ($60\%$), indicating that these instances better align with the capabilities of the participants.

As a next step, we investigate the effect of AI delegation on the overall team performance. To evaluate whether AI delegation achieves complementary team performance, we determine the human-AI team performance as described in Equation \ref{eq:ctp} (see Section \ref{sec:evaluation_measures}). The combined human-AI team performance is $80.01\%$, which is significantly higher ($p<0.001$, one-sample, one-tailed Wilcoxon signed rank test) than the performance of the AI model on the test set ($75.83\%$) and significantly higher ($p<0.001$, one-tailed Mann-Whitney U test) than the performance of the humans ($67.13\%$) working alone. 

To summarize, H1 is supported. When the AI model delegates instances to the participants, their task performance on these images improves, compared to other participants and the AI model conducting the task alone. Task performance improves for both the delegation and the hidden delegation group. The combined human-AI team performance even surpasses the team members' individual performance given they conducted the task independently.

\begin{table*}[t] 
\caption{Regression results: Direct and indirect effect of AI delegation on task performance and task satisfaction.}
\label{tab:results_regression}
\begin{tabular}{llclcllclclc}
\toprule
Regression Model & \multicolumn{4}{c}{Model I---Direct effect of AI delegation} && \multicolumn{6}{c}{Model II---Indirect effect of AI delegation} \\
\midrule
Variable & \multicolumn{2}{c}{Task performance} & \multicolumn{2}{c}{Task satisfaction} && \multicolumn{2}{c}{Self-efficacy} & \multicolumn{2}{c}{Task performance} & \multicolumn{2}{c}{Task satisfaction} \\
\multicolumn{1}{l}{} &   \multicolumn{1}{c}{\textit{coeff}} & \multicolumn{1}{c}{\textit{se}} &  \multicolumn{1}{c}{\textit{coeff}} & \multicolumn{1}{c}{\textit{se}} && \multicolumn{1}{c}{\textit{coeff}} & \multicolumn{1}{c}{\textit{se}} & \multicolumn{1}{c}{\textit{coeff}} & \multicolumn{1}{c}{\textit{se}} & \multicolumn{1}{c}{\textit{coeff}} & \multicolumn{1}{c}{\textit{se}} \\
\midrule
Intercept & ~12.60$^{***}$ & 1.39 & ~1.86$^{***}$ & 0.39 && ~1.44$^{*}$ & 0.63 & 11.92$^{***}$ & 1.46 & ~1.21$^{***}$ & 0.34 \\
AI Delegation & & & & && & & & & & \\
- \textit{Delegation} & ~3.44$^{***}$ & 0.43 & ~0.34$^{**}$ & 0.12 && ~0.75$^{***}$ & 0.18 & ~2.97$^{***}$ & 0.43 & ~0.05 & 0.10 \\
- \textit{Hidden delegation} & ~3.40$^{***}$ & 0.41 & ~0.30$^{*}$ & 0.12 && ~0.80$^{***}$ & 0.17 & ~2.90$^{***}$ & 0.42 & -0.01 & 0.10 \\
- \textit{Human-alone} (baseline) & & & & && & & & & & \\
Self-efficacy & & && & & \multicolumn{2}{c}{} & ~0.62$^{***}$ & 0.17 & ~0.39$^{***}$ & 0.04 \\
Task experience & -0.03 & 0.29 & ~0.13 & 0.08 && ~0.08 & 0.12 & -0.08 & 0.28 & ~0.10 & 0.06  \\
Algorithm attitude & ~0.59$^{**}$ & 0.22 & ~0.10$^{\dagger}$ & 0.06 && ~0.05 & 0.09 & ~0.55$^{*}$ & 0.21 & ~0.08$^{\dagger}$ & 0.05 \\
Algorithm use & -0.31 & 0.20 & -0.09$^{\dagger}$ & 0.06 && -0.06 & 0.08 & -0.28 & 0.19 & -0.07 & 0.04  \\
Cognitive ability & ~0.09 & 0.18 & ~0.21$^{***}$ & 0.05 && ~0.46$^{***}$ & 0.07 & -0.19 & 0.19 & ~0.03 & 0.04 \\
Education & ~0.15 & 0.21 & ~0.00 & 0.06 && ~0.09 & 0.09 & ~0.09 & 0.20 & -0.03 & 0.05  \\
Age & -0.04$^{**}$ & 0.01 & ~0.00 & 0.00 && ~0.01 & 0.01 & -0.04$^{**}$ & 0.01 & ~0.00 & 0.00  \\
Gender & ~0.22 & 0.37 & -0.08 & 0.10 && ~0.17 & 0.16 & -0.32 & 0.36 & ~0.02 & 0.08  \\
\midrule
R$^{2}$& \multicolumn{2}{c}{0.364} & \multicolumn{2}{c}{0.160} && \multicolumn{2}{c}{0.258} & \multicolumn{2}{c}{0.408} & \multicolumn{2}{c}{0.455} \\
Adj. R$^{2}$ & \multicolumn{2}{c}{0.333} & \multicolumn{2}{c}{0.119} && \multicolumn{2}{c}{0.222} & \multicolumn{2}{c}{0.376} & \multicolumn{2}{c}{0.426} \\
MSE & \multicolumn{2}{c}{5.729} & \multicolumn{2}{c}{0.445} && \multicolumn{2}{c}{1.020} & \multicolumn{2}{c}{5.361} & \multicolumn{2}{c}{0.290} \\
F(df) & \multicolumn{2}{c}{11.817$^{***}$ (9,186)} & \multicolumn{2}{c}{3.925$^{***}$ (9,186)} && \multicolumn{2}{c}{7.166$^{***}$ (9,186)} & \multicolumn{2}{c}{12.747$^{***}$(10,185)} & \multicolumn{2}{c}{15.447$^{***}$(10,185)} \\
\bottomrule
\multicolumn{12}{l}{Note: $^{***}$ $p<0.001$; $^{**}$ $p<0.01$; $^{*}$ $p<0.05$; $^{\dagger}$ $p<0.10$}	\\
\end{tabular}
\Description{Regression results: Direct and indirect effect of AI delegation on task performance and task satisfaction.}
\end{table*}

\subsection{Effect of AI Delegation on Task Satisfaction}

\begin{figure}[b]
  \centering
  \includegraphics[width=\linewidth]{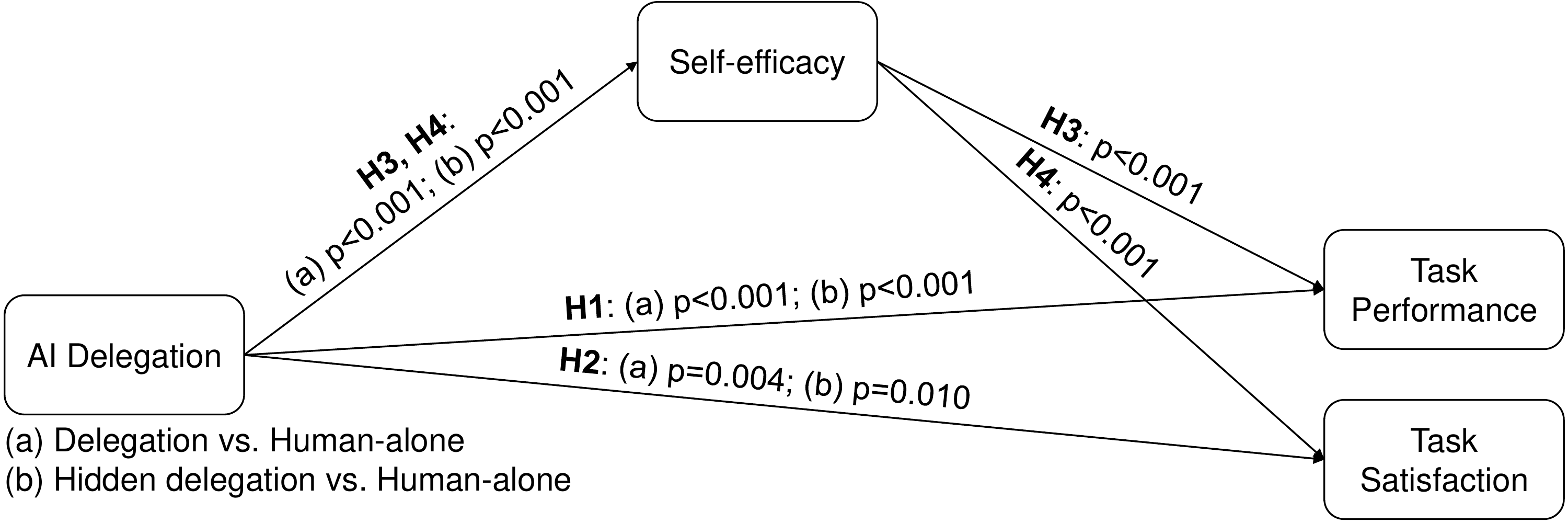}
  \caption{Overview of the direct and indirect effect of AI delegation on task performance and task satisfaction.}
  \label{Fig:results_mediation}
  \Description{Overview of the direct and indirect effect of AI delegation on task performance and task satisfaction.}
\end{figure}

The overall regression model---testing for the direct effect of AI delegation on task satisfaction---is significant, $F(9,186)=3.925, R^2=0.160, p<0.001$. 
Participants in both the delegation and hidden delegation group yield higher levels of task satisfaction ($Mean=3.65, SD=0.66, p<0.004$ and $Mean=3.62, SD=0.73, p<0.010$, respectively), compared to humans working alone ($Mean=3.35, SD=0.70$). A Tukey post hoc test reveals no significant difference in task satisfaction between the two delegation groups ($p=0.961$). Figure \ref{Fig:satisfaction} displays the results of all three groups. We observe that higher cognitive abilities strongly improve ($p<0.001$), and a more positive attitude towards algorithms slightly improves task satisfaction ($p<0.094$). Interestingly, participants who use algorithms more often experience slightly lower task satisfaction ($p<0.097$). For an overview of the regression results, see Figure \ref{Fig:results_mediation} and Table \ref{tab:results_regression} (Columns: `Model I---Direct effect of AI delegation', `Task satisfaction'). 

We conclude that H2 is supported. When the AI delegates task instances to the participants, task satisfaction improves compared to participants working alone. Interestingly, participants' task satisfaction improves significantly, regardless of whether the delegation is communicated to them or not. 

\subsection{Mediation of Self-efficacy in Effect of AI Delegation on Task Performance}

The mediation model---testing for the indirect effect of AI delegation on task performance through increased self-efficacy---is significant, $F(10,185)=12.747, R^2=0.408, p<0.001$. Participants in both the delegation and hidden delegation group have higher self-efficacy ($Mean=5.29, SD=0.96,p<0.001$ and $Mean=5.37, SD=1.05, p<0.001$, respectively), compared to participants in the human-alone group ($Mean=4.63, SD=1.25$), see Figure \ref{Fig:self_efficacy}. A Tukey post hoc test reveals no significant difference in self-efficacy between the two delegation groups ($p=0.904$). Besides AI delegation, high levels of cognitive ability increase self-efficacy ($p<0.001$). This increased self-efficacy improves task performance ($p<0.001$). A more positive attitude towards algorithms and a younger age also improve task performance ($p<0.011$ and $p<0.002$, respectively). Results of the mediation analysis are displayed in Figure \ref{Fig:results_mediation} and Table \ref{tab:results_regression} (Columns: `Model II---Indirect effect of AI delegation', `Task performance'). Self-efficacy mediates the effect of AI delegation---for both the delegation and hidden delegation group---on task performance, as the mediation indices show ($\beta=0.47$, $SE=0.18$, $95\%$ CI $[0.15, 0.84]$ and $\beta=0.50$, $SE=0.20$, $95\%$ CI $[0.15, 0.94]$, respectively). 

To summarize, H3 is supported. For participants in the delegation groups---whether communicated or not---self-efficacy increases and improves task performance compared to participants working alone (see Table \ref{tab:results_regression}). This means that their self-efficacy increases, regardless of whether they are informed about the delegation or not.

\subsection{Mediation of Self-efficacy in Effect of AI Delegation on Task Satisfaction} 

The mediation model---testing for the indirect effect of AI delegation on task satisfaction through increased self-efficacy---is significant, $F(10,185)=15.447, R^2=0.455, p<0.001$. We already know that AI delegation increases participants' self-efficacy compared to when humans work alone. Besides task performance, this increased self-efficacy also improves task satisfaction ($p<0.001$). A more positive attitude towards algorithms marginally improves task satisfaction ($p<0.097$). Results of the mediation analysis are displayed in Figure \ref{Fig:results_mediation} and Table \ref{tab:results_regression} (Columns: `Model II---Indirect effect of AI delegation', `Task satisfaction'). Self-efficacy mediates the effect of AI delegation---for both the delegation and hidden delegation group---on task satisfaction, as the mediation indices show ($\beta=0.29$, $SE=0.08$, $95\%$ CI $[0.15, 0.46]$ and $\beta=0.31$, $SE=0.08$, $95\%$ CI $[0.17, 0.47]$, respectively).

We conclude that H4 is supported. Participants in the delegation group show increased self-efficacy and thereby improved task satisfaction compared to participants working alone (see Table \ref{tab:results_regression}). Participants' self-efficacy increases, regardless of whether the delegation is communicated to them or not.

\section{Discussion}
\label{sec:discussion}
The main goal of this study was to investigate how and why AI delegation affects human task performance and task satisfaction. We developed a research model inspired by organizational behavior research and tested it using an image classification task. AI delegation refers to both the actual act of allocating task instances and the communication of the delegation to the human team members.

Our results demonstrate that AI delegation improves human task performance, regardless of whether humans know about the delegation taking place. Awareness about delegation neither boosts nor reduces human task performance in the current study. Humans receive exactly those images that match their skills. When working together, this effect results in complementary team performance---i.e., the combined human-AI team performance surpasses both human and AI model performance compared to either conducting the task alone. 

In addition to task performance, we were also interested in the impact of AI delegation on human task satisfaction. Task or, more generally, job satisfaction is critical because it predicts employee well-being \cite{Gerdenitsch2017}, productivity \cite{sadeghian2022artificial}, and commitment to the organization \cite{shore1989job}. Our study shows that AI delegation increases task satisfaction, regardless of whether humans know that an AI model delegates instances of the task. As previously stated, knowing about the AI model that takes on the role of a manager leaves task satisfaction unaffected. 

To understand why the observed effects of AI delegation on task performance and task satisfaction occur, the proposed behavioral model allows us to analyze a possible underlying mechanism, i.e., self-efficacy. We find that the effects of AI delegation on task performance and task satisfaction are driven by an increase in humans' self-efficacy. In other words, humans are more confident in their own ability to complete the task when it is composed by the AI model. As a result, they perform better and are more satisfied with the task. While self-efficacy partially mediates the effect of AI delegation on human task performance, it fully mediates the effect of AI delegation on task satisfaction.

\textit{\textbf{Choice of the Task.}}
The following factors determined how we selected our experimental setting and task: First, AI delegation is usually useful in domains where many individual decisions need to be made. Moreover, the AI model has to be able to conduct the task independently. Otherwise, the allocation of instances between the AI model and humans is not possible. We chose our task with these prerequisites in mind and selected image classification as a test bed to evaluate how and why AI delegation influences humans. We believe that image classification is a suitable delegation task since there are many real-world situations where humans need to classify many individual images. Tasks can range from low-stakes tasks, such as animal classification \cite{Bondi2022,8354227}, to high-stakes tasks, such as cancer detection \cite{hekler2019superior}. Additionally, image classification is a task where prior research has shown that humans and AI have complementary strengths and, thus, the potential to reach complementary team performance exists \cite{geirhos2021partial}.

\textit{\textbf{Implications for Human-AI Collaboration.}}
AI delegation, as a special case of human-AI collaboration, has the potential to reduce human effort in tedious tasks and improve human and overall team performance. Prior work has focused on delegation algorithms in user studies that do not learn both the capabilities of humans and the AI model \cite{Bondi2022,fuegener2022cognitive}. Moreover, these studies do not consider humans' perceptions when an AI model manages the delegation of instances. The current study does not only confirm the benefits of AI delegation in general, it also demonstrates the advantages when the capabilities of both team members are taken into account. Furthermore, it enables us to provide insights into humans' perception of the AI model as the ``manager'', distributing task instances between team members. Our study identifies self-efficacy as an underlying mechanism for the effect of AI delegation on task performance and task satisfaction. Hence, managers could consider applying AI delegation to yield higher levels of performance and employee satisfaction. Interestingly, communicating the AI delegation did not further affect self-efficacy, task performance, and task satisfaction. We can conclude that the modified nature of the task through AI delegation was responsible for the increases in task performance and task-related perceptions. Whether modified tasks increase self-efficacy in general and are perceived as satisfying may depend on humans' preferences, personality, and task context. Some people like tasks that are challenging for them, while others prefer more trivial tasks. 

\textit{\textbf{Implications for Algorithmic Management.}}
In the following, we discuss possible implications for the design of ``AI managers''---a special case of algorithmic management \cite{lee2015working}. Algorithmic management can be understood as transferring managerial functions to algorithms, which is, for example, a central element of the gig economy \cite{benlian2022algorithmic,lee2015working,noponen2019impact}. The gig economy focuses on tasks with many repetitions, such as language translation or image classification. Gig economy platform providers such as Uber organize the matching and delegation of instances based on algorithms. Usually, these algorithms distribute instances to different employees \cite{benlian2022algorithmic}. The AI delegation presented here differentiates from this setting by fulfilling both the role of a manager and an employee. This could open up new scaling potential in the gig economy. For example, digital services could be processed either by humans or algorithms, depending on different criteria, e.g., the urgency of task completion, special task requirements, or the availability of human service providers.

Algorithmic management is usually seen as a double-sided sword. On the one hand, it may lead to efficiency and even performance gains, which is important for the scalability of platform business models \cite{benlian2022algorithmic}. The current research also shows improved perceptions, i.e., self-efficacy and task satisfaction. On the other hand, algorithmic management may induce uncertainty and discomfort among employees \cite{benlian2022algorithmic}. For example, a study on Uber drivers shows that some drivers associate negative feelings with working ``for'' an algorithm \cite{mohlmann2019people}. Future research should examine further when and why people perceive algorithmic management as positive or negative.

To summarize, we wanted to illustrate the existing potential for implementing ``AI managers'' in human-AI collaborations. AI delegation can yield higher task performance and task satisfaction through increased feelings of competence in completing the task.

\textit{\textbf{Limitations.}}
We do not observe any effect of communicating that delegation takes place through an AI model. Future research should investigate other forms of communication and task settings to verify the robustness of this finding. For example, we suggest including explanations for the delegation rationale, e.g., why and in which cases the AI delegates task instances. Furthermore, our current experimental design did not examine human delegation. Previous research has shown that humans generally have difficulty correctly assessing their own abilities compared to an AI \cite{fuegener2022cognitive}. Hence, it is likely that human delegation would result in lower performance levels. Moreover, we conducted our study with non-experts drawing upon a non-specialized task. In environments that require expert knowledge, AI delegation may have different effects on human behavior, such as a greater desire for agency or transparency of AI decisions \cite{vossing2022designing}. 

\textit{\textbf{Future Work.}}
The potential of AI delegation as a lever to improve task-related outcomes opens up several opportunities for future research. People who identify with their work may perceive AI delegation differently. For example, if an employee sees strong meaning in performing a task, AI delegation could be seen as something negative that takes away the desired work. On the other hand, if the work is perceived as tedious, AI delegation could be seen as positive. Whether AI delegation is perceived as positive or negative could also vary greatly from person to person. Future work could examine personality traits of people who are more willing to participate in AI delegation to identify differences in people's reactions to AI delegation. In addition, algorithmic opacity, which refers to the transparency of the delegation algorithm, is a major issue in the algorithmic management literature. We chose to communicate the delegation decision to the human without explaining the rationale for that decision. Research in other areas shows that additional information, e.g., the confidence of the algorithm or explanations for a particular decision, can help improve decision-making performance \cite{Bansal2021,alufaisan_does_2020}. We propose investigating whether additional information, e.g., information indicating why or in which cases task instances are delegated, may affect various task outcomes. Lastly, it may be interesting to test whether task performance and task satisfaction can be further improved by personalized delegation or design features lowering the psychological distance.

\section{Conclusion}
\label{sec:conclusion} 
 
This work studies AI delegation as a special form of human-AI collaboration from a human-centered perspective. We propose a behavioral model that allows us to investigate not only the effect of AI delegation on human task performance and task satisfaction but also to understand why the proposed effects occur. Our findings show that AI delegation improves human task performance and task satisfaction while increases in humans' self-efficacy to complete the task explain these positive effects. The question arises whether ``humans managed by AI models'' can be a suitable form of collaboration for particular workplace settings.

\bibliographystyle{ACM-Reference-Format}
\bibliography{sample-base}


\begin{thebibliography}{75}


\ifx \showCODEN    \undefined \def \showCODEN     #1{\unskip}     \fi
\ifx \showDOI      \undefined \def \showDOI       #1{#1}\fi
\ifx \showISBNx    \undefined \def \showISBNx     #1{\unskip}     \fi
\ifx \showISBNxiii \undefined \def \showISBNxiii  #1{\unskip}     \fi
\ifx \showISSN     \undefined \def \showISSN      #1{\unskip}     \fi
\ifx \showLCCN     \undefined \def \showLCCN      #1{\unskip}     \fi
\ifx \shownote     \undefined \def \shownote      #1{#1}          \fi
\ifx \showarticletitle \undefined \def \showarticletitle #1{#1}   \fi
\ifx \showURL      \undefined \def \showURL       {\relax}        \fi
\providecommand\bibfield[2]{#2}
\providecommand\bibinfo[2]{#2}
\providecommand\natexlab[1]{#1}
\providecommand\showeprint[2][]{arXiv:#2}

\bibitem[Adadi and Berrada(2018)]%
        {adadi_peeking_2018}
\bibfield{author}{\bibinfo{person}{Amina Adadi} {and} \bibinfo{person}{Mohammed
  Berrada}.} \bibinfo{year}{2018}\natexlab{}.
\newblock \showarticletitle{Peeking inside the black-box: a survey on
  explainable artificial intelligence (XAI)}.
\newblock \bibinfo{journal}{\emph{IEEE Access}}  \bibinfo{volume}{6}
  (\bibinfo{year}{2018}), \bibinfo{pages}{52138--52160}.
\newblock


\bibitem[Afzal et~al\mbox{.}(2019)]%
        {Afzal2019}
\bibfield{author}{\bibinfo{person}{Sadia Afzal}, \bibinfo{person}{Muhammad
  Arshad}, \bibinfo{person}{Sharjeel Saleem}, {and} \bibinfo{person}{Omer
  Farooq}.} \bibinfo{year}{2019}\natexlab{}.
\newblock \showarticletitle{The impact of perceived supervisor support on
  employees’ turnover intention and task performance: mediation of
  self-efficacy}.
\newblock \bibinfo{journal}{\emph{Journal of Management Development}}
  \bibinfo{volume}{38}, \bibinfo{number}{5} (\bibinfo{year}{2019}),
  \bibinfo{pages}{369--382}.
\newblock


\bibitem[Al-Jammal et~al\mbox{.}(2015)]%
        {al2015impact}
\bibfield{author}{\bibinfo{person}{Hamdan~Rasheed Al-Jammal},
  \bibinfo{person}{Akif~Lutfi Al-Khasawneh}, {and}
  \bibinfo{person}{Mohammad~Hasan Hamadat}.} \bibinfo{year}{2015}\natexlab{}.
\newblock \showarticletitle{The impact of the delegation of authority on
  employees’ performance at great Irbid municipality: case study}.
\newblock \bibinfo{journal}{\emph{International Journal of Human Resource
  Studies}} \bibinfo{volume}{5}, \bibinfo{number}{3} (\bibinfo{year}{2015}),
  \bibinfo{pages}{48--69}.
\newblock


\bibitem[Alufaisan et~al\mbox{.}(2021)]%
        {alufaisan_does_2020}
\bibfield{author}{\bibinfo{person}{Yasmeen Alufaisan}, \bibinfo{person}{Laura~R
  Marusich}, \bibinfo{person}{Jonathan~Z Bakdash}, \bibinfo{person}{Yan Zhou},
  {and} \bibinfo{person}{Murat Kantarcioglu}.} \bibinfo{year}{2021}\natexlab{}.
\newblock \showarticletitle{Does explainable artificial intelligence improve
  human decision-making?}. In \bibinfo{booktitle}{\emph{Proceedings of the AAAI
  Conference on Artificial Intelligence}}. \bibinfo{pages}{6618--6626}.
\newblock


\bibitem[Bandura(1977)]%
        {Bandura1977}
\bibfield{author}{\bibinfo{person}{Albert Bandura}.}
  \bibinfo{year}{1977}\natexlab{}.
\newblock \showarticletitle{Self-efficacy: toward a unifying theory of
  behavioral change}.
\newblock \bibinfo{journal}{\emph{Psychological Review}} \bibinfo{volume}{84},
  \bibinfo{number}{2} (\bibinfo{year}{1977}), \bibinfo{pages}{191--215}.
\newblock


\bibitem[Bansal et~al\mbox{.}(2021)]%
        {Bansal2021}
\bibfield{author}{\bibinfo{person}{Gagan Bansal}, \bibinfo{person}{Tongshuang
  Wu}, \bibinfo{person}{Joyce Zhou}, \bibinfo{person}{Raymond Fok},
  \bibinfo{person}{Besmira Nushi}, \bibinfo{person}{Ece Kamar},
  \bibinfo{person}{Marco~Tulio Ribeiro}, {and} \bibinfo{person}{Daniel Weld}.}
  \bibinfo{year}{2021}\natexlab{}.
\newblock \showarticletitle{Does the whole exceed its parts? The effect of AI
  explanations on complementary team performance}. In
  \bibinfo{booktitle}{\emph{Proceedings of the 2021 CHI Conference on Human
  Factors in Computing Systems}}. \bibinfo{pages}{1--16}.
\newblock


\bibitem[Bass and Stogdill(1990)]%
        {bass1990bass}
\bibfield{author}{\bibinfo{person}{Bernard~M Bass} {and}
  \bibinfo{person}{Ralph~Melvin Stogdill}.} \bibinfo{year}{1990}\natexlab{}.
\newblock \bibinfo{booktitle}{\emph{Bass \& Stogdill's Handbook of Leadership:
  Theory, Research, and Managerial Applications}}.
\newblock \bibinfo{publisher}{Simon and Schuster}.
\newblock


\bibitem[Baumeister et~al\mbox{.}(2002)]%
        {baumeister2002}
\bibfield{author}{\bibinfo{person}{Roy~F Baumeister},
  \bibinfo{person}{Kathleen~D Vohs}, {et~al\mbox{.}}}
  \bibinfo{year}{2002}\natexlab{}.
\newblock \showarticletitle{The pursuit of meaningfulness in life}.
\newblock \bibinfo{journal}{\emph{Handbook of Positive Psychology}}
  \bibinfo{volume}{1} (\bibinfo{year}{2002}), \bibinfo{pages}{608--618}.
\newblock


\bibitem[Benlian et~al\mbox{.}(2022)]%
        {benlian2022algorithmic}
\bibfield{author}{\bibinfo{person}{Alexander Benlian}, \bibinfo{person}{Martin
  Wiener}, \bibinfo{person}{W~Alec Cram}, \bibinfo{person}{Hanna Krasnova},
  \bibinfo{person}{Alexander Maedche}, \bibinfo{person}{Mareike M{\"o}hlmann},
  \bibinfo{person}{Jan Recker}, {and} \bibinfo{person}{Ulrich Remus}.}
  \bibinfo{year}{2022}\natexlab{}.
\newblock \showarticletitle{Algorithmic management: bright and dark sides,
  practical implications, and research opportunities}.
\newblock \bibinfo{journal}{\emph{Business \& Information Systems Engineering}}
  \bibinfo{volume}{64}, \bibinfo{number}{6} (\bibinfo{year}{2022}),
  \bibinfo{pages}{825--839}.
\newblock


\bibitem[Bilal et~al\mbox{.}(2022)]%
        {bilal2022ai}
\bibfield{author}{\bibinfo{person}{Mohsin Bilal}, \bibinfo{person}{Yee~Wah
  Tsang}, \bibinfo{person}{Mahmoud Ali}, \bibinfo{person}{Simon Graham},
  \bibinfo{person}{Emily Hero}, \bibinfo{person}{Noorul Wahab},
  \bibinfo{person}{Katherine Dodd}, \bibinfo{person}{Harvir Sahota},
  \bibinfo{person}{Wenqi Lu}, \bibinfo{person}{Mostafa Jahanifar},
  {et~al\mbox{.}}} \bibinfo{year}{2022}\natexlab{}.
\newblock \showarticletitle{AI based pre-screening of large bowel cancer via
  weakly supervised learning of colorectal biopsy histology images}.
\newblock \bibinfo{journal}{\emph{medRxiv}} (\bibinfo{year}{2022}).
\newblock


\bibitem[Bondi et~al\mbox{.}(2022)]%
        {Bondi2022}
\bibfield{author}{\bibinfo{person}{Elizabeth Bondi}, \bibinfo{person}{Raphael
  Koster}, \bibinfo{person}{Hannah Sheahan}, \bibinfo{person}{Martin Chadwick},
  \bibinfo{person}{Yoram Bachrach}, \bibinfo{person}{Taylan Cemgil},
  \bibinfo{person}{Ulrich Paquet}, {and} \bibinfo{person}{Krishnamurthy
  Dvijotham}.} \bibinfo{year}{2022}\natexlab{}.
\newblock \showarticletitle{Role of human-AI interaction in selective
  prediction}. In \bibinfo{booktitle}{\emph{Proceedings of the AAAI Conference
  on Artificial Intelligence}}. \bibinfo{pages}{5286--5294}.
\newblock


\bibitem[Brown and Sandholm(2019)]%
        {brown2019superhuman}
\bibfield{author}{\bibinfo{person}{Noam Brown} {and} \bibinfo{person}{Tuomas
  Sandholm}.} \bibinfo{year}{2019}\natexlab{}.
\newblock \showarticletitle{Superhuman AI for multiplayer poker}.
\newblock \bibinfo{journal}{\emph{Science}} \bibinfo{volume}{365},
  \bibinfo{number}{6456} (\bibinfo{year}{2019}), \bibinfo{pages}{885--890}.
\newblock


\bibitem[Bu{\c{c}}inca et~al\mbox{.}(2020)]%
        {bucinca_proxy_2020}
\bibfield{author}{\bibinfo{person}{Zana Bu{\c{c}}inca}, \bibinfo{person}{Phoebe
  Lin}, \bibinfo{person}{Krzysztof~Z Gajos}, {and} \bibinfo{person}{Elena~L
  Glassman}.} \bibinfo{year}{2020}\natexlab{}.
\newblock \showarticletitle{Proxy tasks and subjective measures can be
  misleading in evaluating explainable AI systems}. In
  \bibinfo{booktitle}{\emph{Proceedings of the 25th International Conference on
  Intelligent User Interfaces}}. \bibinfo{pages}{454--464}.
\newblock


\bibitem[Bu{\c{c}}inca et~al\mbox{.}(2021)]%
        {bucinca_trust_2021}
\bibfield{author}{\bibinfo{person}{Zana Bu{\c{c}}inca},
  \bibinfo{person}{Maja~Barbara Malaya}, {and} \bibinfo{person}{Krzysztof~Z
  Gajos}.} \bibinfo{year}{2021}\natexlab{}.
\newblock \showarticletitle{To trust or to think: cognitive forcing functions
  can reduce overreliance on AI in AI-assisted decision-making}.
\newblock \bibinfo{journal}{\emph{Proceedings of the ACM on Human-Computer
  Interaction}} \bibinfo{volume}{5}, \bibinfo{number}{CSCW1}
  (\bibinfo{year}{2021}), \bibinfo{pages}{1--21}.
\newblock


\bibitem[Carton et~al\mbox{.}(2020)]%
        {carton_feature-based_2020}
\bibfield{author}{\bibinfo{person}{Samuel Carton}, \bibinfo{person}{Qiaozhu
  Mei}, {and} \bibinfo{person}{Paul Resnick}.} \bibinfo{year}{2020}\natexlab{}.
\newblock \showarticletitle{Feature-based explanations don't help people detect
  misclassifications of online toxicity}. In
  \bibinfo{booktitle}{\emph{Proceedings of the International AAAI Conference on
  Web and Social Media}}. \bibinfo{pages}{95--106}.
\newblock


\bibitem[Desmond et~al\mbox{.}(2021)]%
        {Desmond_2022}
\bibfield{author}{\bibinfo{person}{Michael Desmond}, \bibinfo{person}{Michael
  Muller}, \bibinfo{person}{Zahra Ashktorab}, \bibinfo{person}{Casey Dugan},
  \bibinfo{person}{Evelyn Duesterwald}, \bibinfo{person}{Kristina Brimijoin},
  \bibinfo{person}{Catherine Finegan-Dollak}, \bibinfo{person}{Michelle
  Brachman}, \bibinfo{person}{Aabhas Sharma}, \bibinfo{person}{Narendra~Nath
  Joshi}, {and} \bibinfo{person}{Qian Pan}.} \bibinfo{year}{2021}\natexlab{}.
\newblock \showarticletitle{Increasing the speed and accuracy of data labeling
  through an AI assisted interface}. In \bibinfo{booktitle}{\emph{Proceedings
  of the 26th International Conference on Intelligent User Interfaces}}.
  \bibinfo{pages}{392–401}.
\newblock


\bibitem[Dietvorst et~al\mbox{.}(2015)]%
        {dietvorst2015algorithm}
\bibfield{author}{\bibinfo{person}{Berkeley~J Dietvorst},
  \bibinfo{person}{Joseph~P Simmons}, {and} \bibinfo{person}{Cade Massey}.}
  \bibinfo{year}{2015}\natexlab{}.
\newblock \showarticletitle{Algorithm aversion: people erroneously avoid
  algorithms after seeing them err.}
\newblock \bibinfo{journal}{\emph{Journal of Experimental Psychology: General}}
  \bibinfo{volume}{144}, \bibinfo{number}{1} (\bibinfo{year}{2015}),
  \bibinfo{pages}{114}.
\newblock


\bibitem[Esteva et~al\mbox{.}(2017)]%
        {esteva2017dermatologist}
\bibfield{author}{\bibinfo{person}{Andre Esteva}, \bibinfo{person}{Brett
  Kuprel}, \bibinfo{person}{Roberto~A Novoa}, \bibinfo{person}{Justin Ko},
  \bibinfo{person}{Susan~M Swetter}, \bibinfo{person}{Helen~M Blau}, {and}
  \bibinfo{person}{Sebastian Thrun}.} \bibinfo{year}{2017}\natexlab{}.
\newblock \showarticletitle{Dermatologist-level classification of skin cancer
  with deep neural networks}.
\newblock \bibinfo{journal}{\emph{Nature}} \bibinfo{volume}{542},
  \bibinfo{number}{7639} (\bibinfo{year}{2017}), \bibinfo{pages}{115--118}.
\newblock


\bibitem[Farrow et~al\mbox{.}(1980)]%
        {farrow1980comparison}
\bibfield{author}{\bibinfo{person}{Dana~L Farrow}, \bibinfo{person}{Enzo~R
  Valenzi}, {and} \bibinfo{person}{Bernard~M Bass}.}
  \bibinfo{year}{1980}\natexlab{}.
\newblock \showarticletitle{A comparison of leadership and situational
  characteristics within profit and non-profit organizations.}
\newblock \bibinfo{journal}{\emph{Academy of Management Proceedings}}
  \bibinfo{volume}{1980}, \bibinfo{number}{1} (\bibinfo{year}{1980}),
  \bibinfo{pages}{334--338}.
\newblock


\bibitem[Faul et~al\mbox{.}(2007)]%
        {faul2007g}
\bibfield{author}{\bibinfo{person}{Franz Faul}, \bibinfo{person}{Edgar
  Erdfelder}, \bibinfo{person}{Albert-Georg Lang}, {and} \bibinfo{person}{Axel
  Buchner}.} \bibinfo{year}{2007}\natexlab{}.
\newblock \showarticletitle{G* Power 3: a flexible statistical power analysis
  program for the social, behavioral, and biomedical sciences}.
\newblock \bibinfo{journal}{\emph{Behavior Research Methods}}
  \bibinfo{volume}{39}, \bibinfo{number}{2} (\bibinfo{year}{2007}),
  \bibinfo{pages}{175--191}.
\newblock


\bibitem[Feng and Boyd-Graber(2019)]%
        {feng2019can}
\bibfield{author}{\bibinfo{person}{Shi Feng} {and} \bibinfo{person}{Jordan
  Boyd-Graber}.} \bibinfo{year}{2019}\natexlab{}.
\newblock \showarticletitle{What can AI do for me? Evaluating machine learning
  interpretations in cooperative play}. In
  \bibinfo{booktitle}{\emph{Proceedings of the 24th International Conference on
  Intelligent User Interfaces}}. \bibinfo{pages}{229--239}.
\newblock


\bibitem[F\"{u}gener et~al\mbox{.}(2022)]%
        {fuegener2022cognitive}
\bibfield{author}{\bibinfo{person}{Andreas F\"{u}gener},
  \bibinfo{person}{J\"{o}rn Grahl}, \bibinfo{person}{Alok Gupta}, {and}
  \bibinfo{person}{Wolfgang Ketter}.} \bibinfo{year}{2022}\natexlab{}.
\newblock \showarticletitle{Cognitive challenges in human--artificial
  intelligence collaboration: investigating the path toward productive
  delegation}.
\newblock \bibinfo{journal}{\emph{Information Systems Research}}
  \bibinfo{volume}{33}, \bibinfo{number}{2} (\bibinfo{year}{2022}),
  \bibinfo{pages}{678--696}.
\newblock


\bibitem[Gecas(1982)]%
        {gecas1982}
\bibfield{author}{\bibinfo{person}{Viktor Gecas}.}
  \bibinfo{year}{1982}\natexlab{}.
\newblock \showarticletitle{The self-concept}.
\newblock \bibinfo{journal}{\emph{Annual Review of Sociology}}
  \bibinfo{volume}{8}, \bibinfo{number}{1} (\bibinfo{year}{1982}),
  \bibinfo{pages}{1--33}.
\newblock


\bibitem[Geirhos et~al\mbox{.}(2021)]%
        {geirhos2021partial}
\bibfield{author}{\bibinfo{person}{Robert Geirhos}, \bibinfo{person}{Kantharaju
  Narayanappa}, \bibinfo{person}{Benjamin Mitzkus}, \bibinfo{person}{Tizian
  Thieringer}, \bibinfo{person}{Matthias Bethge}, \bibinfo{person}{Felix~A
  Wichmann}, {and} \bibinfo{person}{Wieland Brendel}.}
  \bibinfo{year}{2021}\natexlab{}.
\newblock \showarticletitle{Partial success in closing the gap between human
  and machine vision}. In \bibinfo{booktitle}{\emph{Advances in Neural
  Information Processing Systems}}. \bibinfo{pages}{23885--23899}.
\newblock


\bibitem[Gerdenitsch(2017)]%
        {Gerdenitsch2017}
\bibfield{author}{\bibinfo{person}{Cornelia Gerdenitsch}.}
  \bibinfo{year}{2017}\natexlab{}.
\newblock \showarticletitle{New ways of working and satisfaction of
  psychological needs}.
\newblock In \bibinfo{booktitle}{\emph{Job Demands in a Changing World of
  Work}}. \bibinfo{pages}{91--109}.
\newblock


\bibitem[Gist(1987)]%
        {gist1987self}
\bibfield{author}{\bibinfo{person}{Marilyn~E Gist}.}
  \bibinfo{year}{1987}\natexlab{}.
\newblock \showarticletitle{Self-efficacy: implications for organizational
  behavior and human resource management}.
\newblock \bibinfo{journal}{\emph{Academy of Management Review}}
  \bibinfo{volume}{12}, \bibinfo{number}{3} (\bibinfo{year}{1987}),
  \bibinfo{pages}{472--485}.
\newblock


\bibitem[Gulshan et~al\mbox{.}(2019)]%
        {gulshan2019performance}
\bibfield{author}{\bibinfo{person}{Varun Gulshan}, \bibinfo{person}{Renu~P
  Rajan}, \bibinfo{person}{Kasumi Widner}, \bibinfo{person}{Derek Wu},
  \bibinfo{person}{Peter Wubbels}, \bibinfo{person}{Tyler Rhodes},
  \bibinfo{person}{Kira Whitehouse}, \bibinfo{person}{Marc Coram},
  \bibinfo{person}{Greg Corrado}, \bibinfo{person}{Kim Ramasamy},
  \bibinfo{person}{Rajiv Raman}, \bibinfo{person}{Lily Peng}, {and}
  \bibinfo{person}{Dale~R Webster}.} \bibinfo{year}{2019}\natexlab{}.
\newblock \showarticletitle{{Performance of a deep-learning algorithm vs manual
  grading for detecting diabetic retinopathy in India}}.
\newblock \bibinfo{journal}{\emph{JAMA Ophthalmology}} \bibinfo{volume}{137},
  \bibinfo{number}{9} (\bibinfo{year}{2019}), \bibinfo{pages}{987--993}.
\newblock


\bibitem[Hayes(2017)]%
        {Hayes2017IntroductionApproach}
\bibfield{author}{\bibinfo{person}{Andrew~F Hayes}.}
  \bibinfo{year}{2017}\natexlab{}.
\newblock \bibinfo{booktitle}{\emph{Introduction to Mediation, Moderation, and
  Conditional Process Analysis: A Regression-based Approach}}.
\newblock \bibinfo{publisher}{Guilford publications}.
\newblock


\bibitem[He et~al\mbox{.}(2015)]%
        {he2015delving}
\bibfield{author}{\bibinfo{person}{Kaiming He}, \bibinfo{person}{Xiangyu
  Zhang}, \bibinfo{person}{Shaoqing Ren}, {and} \bibinfo{person}{Jian Sun}.}
  \bibinfo{year}{2015}\natexlab{}.
\newblock \showarticletitle{Delving deep into rectifiers: surpassing
  human-level performance on imagenet classification}. In
  \bibinfo{booktitle}{\emph{Proceedings of the IEEE International Conference on
  Computer Vision}}. \bibinfo{pages}{1026--1034}.
\newblock


\bibitem[Hekler et~al\mbox{.}(2019)]%
        {hekler2019superior}
\bibfield{author}{\bibinfo{person}{Achim Hekler}, \bibinfo{person}{Jochen~S
  Utikal}, \bibinfo{person}{Alexander~H Enk}, \bibinfo{person}{Axel Hauschild},
  \bibinfo{person}{Michael Weichenthal}, \bibinfo{person}{Roman~C Maron},
  \bibinfo{person}{Carola Berking}, \bibinfo{person}{Sebastian Haferkamp},
  \bibinfo{person}{Joachim Klode}, \bibinfo{person}{Dirk Schadendorf},
  {et~al\mbox{.}}} \bibinfo{year}{2019}\natexlab{}.
\newblock \showarticletitle{Superior skin cancer classification by the
  combination of human and artificial intelligence}.
\newblock \bibinfo{journal}{\emph{European Journal of Cancer}}
  \bibinfo{volume}{120} (\bibinfo{year}{2019}), \bibinfo{pages}{114--121}.
\newblock


\bibitem[Hemmer et~al\mbox{.}(2022a)]%
        {ijcai_multiple_experts}
\bibfield{author}{\bibinfo{person}{Patrick Hemmer}, \bibinfo{person}{Sebastian
  Schellhammer}, \bibinfo{person}{Michael Vössing}, \bibinfo{person}{Johannes
  Jakubik}, {and} \bibinfo{person}{Gerhard Satzger}.}
  \bibinfo{year}{2022}\natexlab{a}.
\newblock \showarticletitle{Forming effective human-AI teams: building machine
  learning models that complement the capabilities of multiple experts}. In
  \bibinfo{booktitle}{\emph{Proceedings of the Thirty-First International Joint
  Conference on Artificial Intelligence}}. \bibinfo{pages}{2478--2484}.
\newblock


\bibitem[Hemmer et~al\mbox{.}(2022b)]%
        {hemmer2022effect}
\bibfield{author}{\bibinfo{person}{Patrick Hemmer}, \bibinfo{person}{Max
  Schemmer}, \bibinfo{person}{Niklas K{\"u}hl}, \bibinfo{person}{Michael
  V{\"o}ssing}, {and} \bibinfo{person}{Gerhard Satzger}.}
  \bibinfo{year}{2022}\natexlab{b}.
\newblock \showarticletitle{On the effect of information asymmetry in human-AI
  teams}.
\newblock \bibinfo{journal}{\emph{Human-Centered Explainable AI Workshop at the
  2022 CHI Conference on Human Factors in Computing Systems}}
  (\bibinfo{year}{2022}).
\newblock


\bibitem[Hemmer et~al\mbox{.}(2021)]%
        {hemmer2schemmer021}
\bibfield{author}{\bibinfo{person}{Patrick Hemmer}, \bibinfo{person}{Max
  Schemmer}, \bibinfo{person}{Michael V{\"o}ssing}, {and}
  \bibinfo{person}{Niklas K{\"u}hl}.} \bibinfo{year}{2021}\natexlab{}.
\newblock \showarticletitle{Human-AI complementarity in hybrid intelligence
  systems: a structured literature review}. In
  \bibinfo{booktitle}{\emph{Proceedings of the Pacific Asia Conference on
  Information Systems}}.
\newblock


\bibitem[Hofmann and Strickland(1995)]%
        {Hofmann1995}
\bibfield{author}{\bibinfo{person}{David~A Hofmann} {and}
  \bibinfo{person}{Oriel~J Strickland}.} \bibinfo{year}{1995}\natexlab{}.
\newblock \showarticletitle{Task performance and satisfaction: evidence for a
  task- by ego-orientation interaction}.
\newblock \bibinfo{journal}{\emph{Journal of Applied Social Psychology}}
  \bibinfo{volume}{25}, \bibinfo{number}{6} (\bibinfo{year}{1995}),
  \bibinfo{pages}{495--511}.
\newblock


\bibitem[Huang et~al\mbox{.}(2017)]%
        {huang2017}
\bibfield{author}{\bibinfo{person}{Gao Huang}, \bibinfo{person}{Zhuang Liu},
  \bibinfo{person}{Laurens Van Der~Maaten}, {and} \bibinfo{person}{Kilian~Q
  Weinberger}.} \bibinfo{year}{2017}\natexlab{}.
\newblock \showarticletitle{Densely connected convolutional networks}. In
  \bibinfo{booktitle}{\emph{Proceedings of the IEEE Conference on Computer
  Vision and Pattern Recognition}}. \bibinfo{pages}{4700--4708}.
\newblock


\bibitem[Irvin et~al\mbox{.}(2019)]%
        {irvin2019chexpert}
\bibfield{author}{\bibinfo{person}{Jeremy Irvin}, \bibinfo{person}{Pranav
  Rajpurkar}, \bibinfo{person}{Michael Ko}, \bibinfo{person}{Yifan Yu},
  \bibinfo{person}{Silviana Ciurea-Ilcus}, \bibinfo{person}{Chris Chute},
  \bibinfo{person}{Henrik Marklund}, \bibinfo{person}{Behzad Haghgoo},
  \bibinfo{person}{Robyn Ball}, \bibinfo{person}{Katie Shpanskaya},
  {et~al\mbox{.}}} \bibinfo{year}{2019}\natexlab{}.
\newblock \showarticletitle{Chexpert: a large chest radiograph dataset with
  uncertainty labels and expert comparison}. In
  \bibinfo{booktitle}{\emph{Proceedings of the AAAI Conference on Artificial
  Intelligence}}. \bibinfo{pages}{590--597}.
\newblock


\bibitem[Jacobs and Roodenburg(2014)]%
        {Jacobs2014}
\bibfield{author}{\bibinfo{person}{Kate~E Jacobs} {and} \bibinfo{person}{John
  Roodenburg}.} \bibinfo{year}{2014}\natexlab{}.
\newblock \showarticletitle{The development and validation of the self-report
  measure of cognitive abilities: a multitrait–multimethod study}.
\newblock \bibinfo{journal}{\emph{Intelligence}}  \bibinfo{volume}{42}
  (\bibinfo{year}{2014}), \bibinfo{pages}{5--21}.
\newblock


\bibitem[Kerrigan et~al\mbox{.}(2021)]%
        {Kerrigan2021}
\bibfield{author}{\bibinfo{person}{Gavin Kerrigan}, \bibinfo{person}{Padhraic
  Smyth}, {and} \bibinfo{person}{Mark Steyvers}.}
  \bibinfo{year}{2021}\natexlab{}.
\newblock \showarticletitle{Combining human predictions with model
  probabilities via confusion matrices and calibration}. In
  \bibinfo{booktitle}{\emph{Advances in Neural Information Processing
  Systems}}. \bibinfo{pages}{4421--4434}.
\newblock


\bibitem[Keswani et~al\mbox{.}(2021)]%
        {keswani2021towards}
\bibfield{author}{\bibinfo{person}{Vijay Keswani}, \bibinfo{person}{Matthew
  Lease}, {and} \bibinfo{person}{Krishnaram Kenthapadi}.}
  \bibinfo{year}{2021}\natexlab{}.
\newblock \showarticletitle{Towards unbiased and accurate deferral to multiple
  experts}. In \bibinfo{booktitle}{\emph{Proceedings of the 2021 AAAI/ACM
  Conference on AI, Ethics, and Society}}. \bibinfo{pages}{154--165}.
\newblock


\bibitem[Lai et~al\mbox{.}(2022)]%
        {Lai2022}
\bibfield{author}{\bibinfo{person}{Vivian Lai}, \bibinfo{person}{Samuel
  Carton}, \bibinfo{person}{Rajat Bhatnagar}, \bibinfo{person}{Q~Vera Liao},
  \bibinfo{person}{Yunfeng Zhang}, {and} \bibinfo{person}{Chenhao Tan}.}
  \bibinfo{year}{2022}\natexlab{}.
\newblock \showarticletitle{Human-AI collaboration via conditional delegation:
  a case study of content moderation}. In \bibinfo{booktitle}{\emph{Proceedings
  of the 2022 CHI Conference on Human Factors in Computing Systems}}.
  \bibinfo{pages}{1--18}.
\newblock


\bibitem[Lai et~al\mbox{.}(2020)]%
        {lai_why_2020}
\bibfield{author}{\bibinfo{person}{Vivian Lai}, \bibinfo{person}{Han Liu},
  {and} \bibinfo{person}{Chenhao Tan}.} \bibinfo{year}{2020}\natexlab{}.
\newblock \showarticletitle{"Why is 'Chicago' deceptive?" Towards building
  model-driven tutorials for humans}. In \bibinfo{booktitle}{\emph{Proceedings
  of the 2020 CHI Conference on Human Factors in Computing Systems}}.
  \bibinfo{pages}{1--13}.
\newblock


\bibitem[Lai and Tan(2019)]%
        {lai_human_2019}
\bibfield{author}{\bibinfo{person}{Vivian Lai} {and} \bibinfo{person}{Chenhao
  Tan}.} \bibinfo{year}{2019}\natexlab{}.
\newblock \showarticletitle{On human predictions with explanations and
  predictions of machine learning models: a case study on deception detection}.
  In \bibinfo{booktitle}{\emph{Proceedings of the Conference on Fairness,
  Accountability, and Transparency}}. \bibinfo{pages}{29--38}.
\newblock


\bibitem[Leana(1986)]%
        {Leana1986}
\bibfield{author}{\bibinfo{person}{Carrie~R Leana}.}
  \bibinfo{year}{1986}\natexlab{}.
\newblock \showarticletitle{Predictors and consequences of delegation}.
\newblock \bibinfo{journal}{\emph{Academy of Management Journal}}
  \bibinfo{volume}{29}, \bibinfo{number}{4} (\bibinfo{year}{1986}),
  \bibinfo{pages}{754--774}.
\newblock


\bibitem[Leana(1987)]%
        {leana1987power}
\bibfield{author}{\bibinfo{person}{Carrie~R Leana}.}
  \bibinfo{year}{1987}\natexlab{}.
\newblock \showarticletitle{Power relinquishment versus power sharing:
  theoretical clarification and empirical comparison of delegation and
  participation.}
\newblock \bibinfo{journal}{\emph{Journal of Applied Psychology}}
  \bibinfo{volume}{72}, \bibinfo{number}{2} (\bibinfo{year}{1987}),
  \bibinfo{pages}{228}.
\newblock


\bibitem[Lee et~al\mbox{.}(2015)]%
        {lee2015working}
\bibfield{author}{\bibinfo{person}{Min~Kyung Lee}, \bibinfo{person}{Daniel
  Kusbit}, \bibinfo{person}{Evan Metsky}, {and} \bibinfo{person}{Laura
  Dabbish}.} \bibinfo{year}{2015}\natexlab{}.
\newblock \showarticletitle{Working with machines: the impact of algorithmic
  and data-driven management on human workers}. In
  \bibinfo{booktitle}{\emph{Proceedings of the 33rd annual ACM Conference on
  Human Factors in Computing Systems}}. \bibinfo{pages}{1603--1612}.
\newblock


\bibitem[Leit{\~a}o et~al\mbox{.}(2022)]%
        {leitao2022human}
\bibfield{author}{\bibinfo{person}{Diogo Leit{\~a}o}, \bibinfo{person}{Pedro
  Saleiro}, \bibinfo{person}{M{\'a}rio~AT Figueiredo}, {and}
  \bibinfo{person}{Pedro Bizarro}.} \bibinfo{year}{2022}\natexlab{}.
\newblock \showarticletitle{Human-AI collaboration in decision-making: beyond
  learning to defer}.
\newblock \bibinfo{journal}{\emph{Workshop on Human-Machine Collaboration and
  Teaming at the International Conference on Machine Learning}}
  (\bibinfo{year}{2022}).
\newblock


\bibitem[Locke et~al\mbox{.}(1984)]%
        {Locke1984}
\bibfield{author}{\bibinfo{person}{Edwin~A Locke}, \bibinfo{person}{Elizabeth
  Frederick}, \bibinfo{person}{Cynthia Lee}, {and} \bibinfo{person}{Philip
  Bobko}.} \bibinfo{year}{1984}\natexlab{}.
\newblock \showarticletitle{Effect of self-efficacy, goals, and task strategies
  on task performance}.
\newblock \bibinfo{journal}{\emph{Journal of Applied Psychology}}
  \bibinfo{volume}{69}, \bibinfo{number}{2} (\bibinfo{year}{1984}),
  \bibinfo{pages}{241--251}.
\newblock


\bibitem[Lunenburg(2011)]%
        {lunenburg2011self}
\bibfield{author}{\bibinfo{person}{Fred~C Lunenburg}.}
  \bibinfo{year}{2011}\natexlab{}.
\newblock \showarticletitle{Self-efficacy in the workplace: implications for
  motivation and performance}.
\newblock \bibinfo{journal}{\emph{International Journal of Management,
  Business, and Administration}} \bibinfo{volume}{14}, \bibinfo{number}{1}
  (\bibinfo{year}{2011}), \bibinfo{pages}{1--6}.
\newblock


\bibitem[M{\"o}hlmann and Henfridsson(2019)]%
        {mohlmann2019people}
\bibfield{author}{\bibinfo{person}{Mareike M{\"o}hlmann} {and}
  \bibinfo{person}{Ola Henfridsson}.} \bibinfo{year}{2019}\natexlab{}.
\newblock \showarticletitle{What people hate about being managed by algorithms,
  according to a study of Uber drivers}.
\newblock \bibinfo{journal}{\emph{Harvard Business Review}}
  \bibinfo{volume}{30} (\bibinfo{year}{2019}), \bibinfo{pages}{1--7}.
\newblock


\bibitem[Mozannar and Sontag(2020)]%
        {mozannar2020consistent}
\bibfield{author}{\bibinfo{person}{Hussein Mozannar} {and}
  \bibinfo{person}{David Sontag}.} \bibinfo{year}{2020}\natexlab{}.
\newblock \showarticletitle{Consistent estimators for learning to defer to an
  expert}. In \bibinfo{booktitle}{\emph{Proceedings of the 37th International
  Conference on Machine Learning}}. \bibinfo{pages}{7076--7087}.
\newblock


\bibitem[Nguyen et~al\mbox{.}(2021)]%
        {nguyen2021effectiveness}
\bibfield{author}{\bibinfo{person}{Giang Nguyen}, \bibinfo{person}{Daeyoung
  Kim}, {and} \bibinfo{person}{Anh Nguyen}.} \bibinfo{year}{2021}\natexlab{}.
\newblock \showarticletitle{The effectiveness of feature attribution methods
  and its correlation with automatic evaluation scores}. In
  \bibinfo{booktitle}{\emph{Advances in Neural Information Processing
  Systems}}. \bibinfo{pages}{26422--26436}.
\newblock


\bibitem[Noponen(2019)]%
        {noponen2019impact}
\bibfield{author}{\bibinfo{person}{Niilo Noponen}.}
  \bibinfo{year}{2019}\natexlab{}.
\newblock \showarticletitle{Impact of artificial intelligence on management}.
\newblock \bibinfo{journal}{\emph{Electronic Journal of Business Ethics and
  Organization Studies}} \bibinfo{volume}{24}, \bibinfo{number}{2}
  (\bibinfo{year}{2019}).
\newblock


\bibitem[Nourani et~al\mbox{.}(2020)]%
        {nourani2020role}
\bibfield{author}{\bibinfo{person}{Mahsan Nourani}, \bibinfo{person}{Joanie
  King}, {and} \bibinfo{person}{Eric Ragan}.} \bibinfo{year}{2020}\natexlab{}.
\newblock \showarticletitle{The role of domain expertise in user trust and the
  impact of first impressions with intelligent systems}. In
  \bibinfo{booktitle}{\emph{Proceedings of the AAAI Conference on Human
  Computation and Crowdsourcing}}. \bibinfo{pages}{112--121}.
\newblock


\bibitem[Okati et~al\mbox{.}(2021)]%
        {okati2021differentiable}
\bibfield{author}{\bibinfo{person}{Nastaran Okati}, \bibinfo{person}{Abir De},
  {and} \bibinfo{person}{Manuel Rodriguez}.} \bibinfo{year}{2021}\natexlab{}.
\newblock \showarticletitle{Differentiable learning under triage}. In
  \bibinfo{booktitle}{\emph{Advances in Neural Information Processing
  Systems}}. \bibinfo{pages}{9140--9151}.
\newblock


\bibitem[Parham et~al\mbox{.}(2018)]%
        {8354227}
\bibfield{author}{\bibinfo{person}{Jason Parham}, \bibinfo{person}{Charles
  Stewart}, \bibinfo{person}{Jonathan Crall}, \bibinfo{person}{Daniel
  Rubenstein}, \bibinfo{person}{Jason Holmberg}, {and} \bibinfo{person}{Tanya
  Berger-Wolf}.} \bibinfo{year}{2018}\natexlab{}.
\newblock \showarticletitle{An animal detection pipeline for identification}.
  In \bibinfo{booktitle}{\emph{2018 IEEE Winter Conference on Applications of
  Computer Vision}}. \bibinfo{pages}{1075--1083}.
\newblock


\bibitem[Raghu et~al\mbox{.}(2019)]%
        {Raghu2019}
\bibfield{author}{\bibinfo{person}{Maithra Raghu}, \bibinfo{person}{Katy
  Blumer}, \bibinfo{person}{Greg Corrado}, \bibinfo{person}{Jon Kleinberg},
  \bibinfo{person}{Ziad Obermeyer}, {and} \bibinfo{person}{Sendhil
  Mullainathan}.} \bibinfo{year}{2019}\natexlab{}.
\newblock \showarticletitle{The algorithmic automation problem: prediction,
  triage, and human effort}.
\newblock \bibinfo{journal}{\emph{arXiv preprint arXiv:1903.12220}}
  (\bibinfo{year}{2019}).
\newblock


\bibitem[Rastogi et~al\mbox{.}(2022)]%
        {Rastogi2022}
\bibfield{author}{\bibinfo{person}{Charvi Rastogi}, \bibinfo{person}{Yunfeng
  Zhang}, \bibinfo{person}{Dennis Wei}, \bibinfo{person}{Kush~R Varshney},
  \bibinfo{person}{Amit Dhurandhar}, {and} \bibinfo{person}{Richard Tomsett}.}
  \bibinfo{year}{2022}\natexlab{}.
\newblock \showarticletitle{Deciding fast and slow: the role of cognitive
  biases in AI-assisted decision-making}.
\newblock \bibinfo{journal}{\emph{Proceedings of the ACM on Human-Computer
  Interaction}} \bibinfo{volume}{6}, \bibinfo{number}{CSCW1}
  (\bibinfo{year}{2022}), \bibinfo{pages}{1--22}.
\newblock


\bibitem[Russakovsky et~al\mbox{.}(2015)]%
        {ILSVRC15}
\bibfield{author}{\bibinfo{person}{Olga Russakovsky}, \bibinfo{person}{Jia
  Deng}, \bibinfo{person}{Hao Su}, \bibinfo{person}{Jonathan Krause},
  \bibinfo{person}{Sanjeev Satheesh}, \bibinfo{person}{Sean Ma},
  \bibinfo{person}{Zhiheng Huang}, \bibinfo{person}{Andrej Karpathy},
  \bibinfo{person}{Aditya Khosla}, \bibinfo{person}{Michael Bernstein},
  \bibinfo{person}{Alexander~C Berg}, {and} \bibinfo{person}{Li Fei-Fei}.}
  \bibinfo{year}{2015}\natexlab{}.
\newblock \showarticletitle{{ImageNet large scale visual recognition
  challenge}}.
\newblock \bibinfo{journal}{\emph{International Journal of Computer Vision}}
  \bibinfo{volume}{115}, \bibinfo{number}{3} (\bibinfo{year}{2015}),
  \bibinfo{pages}{211--252}.
\newblock


\bibitem[Sadeghian and Hassenzahl(2022)]%
        {sadeghian2022artificial}
\bibfield{author}{\bibinfo{person}{Shadan Sadeghian} {and}
  \bibinfo{person}{Marc Hassenzahl}.} \bibinfo{year}{2022}\natexlab{}.
\newblock \showarticletitle{The ”artificial” colleague: evaluation of work
  satisfaction in collaboration with non-human coworkers}. In
  \bibinfo{booktitle}{\emph{Proceedings of the 27th International Conference on
  Intelligent User Interfaces}}. \bibinfo{pages}{27--35}.
\newblock


\bibitem[Schemmer et~al\mbox{.}(2022a)]%
        {schemmer2022should}
\bibfield{author}{\bibinfo{person}{Max Schemmer}, \bibinfo{person}{Patrick
  Hemmer}, \bibinfo{person}{Niklas K{\"u}hl}, \bibinfo{person}{Carina Benz},
  {and} \bibinfo{person}{Gerhard Satzger}.} \bibinfo{year}{2022}\natexlab{a}.
\newblock \showarticletitle{Should I follow AI-based advice? Measuring
  appropriate reliance in human-AI decision-making}.
\newblock \bibinfo{journal}{\emph{Workshop on Trust and Reliance in AI-Human
  Teams at the 2022 CHI Conference on Human Factors in Computing Systems}}
  (\bibinfo{year}{2022}).
\newblock


\bibitem[Schemmer et~al\mbox{.}(2022b)]%
        {Schemmer2022_meta}
\bibfield{author}{\bibinfo{person}{Max Schemmer}, \bibinfo{person}{Patrick
  Hemmer}, \bibinfo{person}{Maximilian Nitsche}, \bibinfo{person}{Niklas
  K\"{u}hl}, {and} \bibinfo{person}{Michael V\"{o}ssing}.}
  \bibinfo{year}{2022}\natexlab{b}.
\newblock \showarticletitle{A meta-analysis of the utility of explainable
  artificial intelligence in human-AI decision-making}. In
  \bibinfo{booktitle}{\emph{Proceedings of the 2022 AAAI/ACM Conference on AI,
  Ethics, and Society}}. \bibinfo{pages}{617–626}.
\newblock


\bibitem[Schriesheim et~al\mbox{.}(1998)]%
        {schriesheim1998delegation}
\bibfield{author}{\bibinfo{person}{Chester~A Schriesheim},
  \bibinfo{person}{Linda~L Neider}, {and} \bibinfo{person}{Terri~A Scandura}.}
  \bibinfo{year}{1998}\natexlab{}.
\newblock \showarticletitle{Delegation and leader-member exchange: main
  effects, moderators, and measurement issues}.
\newblock \bibinfo{journal}{\emph{Academy of Management Journal}}
  \bibinfo{volume}{41}, \bibinfo{number}{3} (\bibinfo{year}{1998}),
  \bibinfo{pages}{298--318}.
\newblock


\bibitem[Shore and Martin(1989)]%
        {shore1989job}
\bibfield{author}{\bibinfo{person}{Lynn~McFarlane Shore} {and}
  \bibinfo{person}{Harry~J Martin}.} \bibinfo{year}{1989}\natexlab{}.
\newblock \showarticletitle{Job satisfaction and organizational commitment in
  relation to work performance and turnover intentions}.
\newblock \bibinfo{journal}{\emph{Human Relations}} \bibinfo{volume}{42},
  \bibinfo{number}{7} (\bibinfo{year}{1989}), \bibinfo{pages}{625--638}.
\newblock


\bibitem[Silver et~al\mbox{.}(2018)]%
        {silver2018general}
\bibfield{author}{\bibinfo{person}{David Silver}, \bibinfo{person}{Thomas
  Hubert}, \bibinfo{person}{Julian Schrittwieser}, \bibinfo{person}{Ioannis
  Antonoglou}, \bibinfo{person}{Matthew Lai}, \bibinfo{person}{Arthur Guez},
  \bibinfo{person}{Marc Lanctot}, \bibinfo{person}{Laurent Sifre},
  \bibinfo{person}{Dharshan Kumaran}, \bibinfo{person}{Thore Graepel},
  {et~al\mbox{.}}} \bibinfo{year}{2018}\natexlab{}.
\newblock \showarticletitle{A general reinforcement learning algorithm that
  masters chess, shogi, and Go through self-play}.
\newblock \bibinfo{journal}{\emph{Science}} \bibinfo{volume}{362},
  \bibinfo{number}{6419} (\bibinfo{year}{2018}), \bibinfo{pages}{1140--1144}.
\newblock


\bibitem[Spreitzer(1995)]%
        {spreitzer1995}
\bibfield{author}{\bibinfo{person}{Gretchen~M Spreitzer}.}
  \bibinfo{year}{1995}\natexlab{}.
\newblock \showarticletitle{Psychological empowerment in the workplace:
  dimensions, measurement, and validation}.
\newblock \bibinfo{journal}{\emph{Academy of Management Journal}}
  \bibinfo{volume}{38}, \bibinfo{number}{5} (\bibinfo{year}{1995}),
  \bibinfo{pages}{1442--1465}.
\newblock


\bibitem[Stajkovic and Luthans(1998)]%
        {stajkovic1998self}
\bibfield{author}{\bibinfo{person}{Alexander~D Stajkovic} {and}
  \bibinfo{person}{Fred Luthans}.} \bibinfo{year}{1998}\natexlab{}.
\newblock \showarticletitle{Self-efficacy and work-related performance: a
  meta-analysis}.
\newblock \bibinfo{journal}{\emph{Psychological Bulletin}}
  \bibinfo{volume}{124}, \bibinfo{number}{2} (\bibinfo{year}{1998}),
  \bibinfo{pages}{240}.
\newblock


\bibitem[Steyvers et~al\mbox{.}(2022)]%
        {Steyvers2022}
\bibfield{author}{\bibinfo{person}{Mark Steyvers}, \bibinfo{person}{Heliodoro
  Tejeda}, \bibinfo{person}{Gavin Kerrigan}, {and} \bibinfo{person}{Padhraic
  Smyth}.} \bibinfo{year}{2022}\natexlab{}.
\newblock \showarticletitle{Bayesian modeling of human–AI complementarity}.
\newblock \bibinfo{journal}{\emph{Proceedings of the National Academy of
  Sciences}} \bibinfo{volume}{119}, \bibinfo{number}{11}
  (\bibinfo{year}{2022}), \bibinfo{pages}{1--7}.
\newblock


\bibitem[Ugoani(2020)]%
        {ugoani2020}
\bibfield{author}{\bibinfo{person}{John Ugoani}.}
  \bibinfo{year}{2020}\natexlab{}.
\newblock \showarticletitle{Effective delegation and its impact on employee
  performance}.
\newblock \bibinfo{journal}{\emph{International Journal of Economics and
  Business Administration}} \bibinfo{volume}{6}, \bibinfo{number}{3}
  (\bibinfo{year}{2020}), \bibinfo{pages}{78--87}.
\newblock


\bibitem[van~der Waa et~al\mbox{.}(2021)]%
        {van_der_waa_evaluating_2021}
\bibfield{author}{\bibinfo{person}{Jasper van~der Waa},
  \bibinfo{person}{Elisabeth Nieuwburg}, \bibinfo{person}{Anita Cremers}, {and}
  \bibinfo{person}{Mark Neerincx}.} \bibinfo{year}{2021}\natexlab{}.
\newblock \showarticletitle{Evaluating XAI: a comparison of rule-based and
  example-based explanations}.
\newblock \bibinfo{journal}{\emph{Artificial Intelligence}}
  \bibinfo{volume}{291} (\bibinfo{year}{2021}), \bibinfo{pages}{103404}.
\newblock


\bibitem[V{\"o}ssing et~al\mbox{.}(2022)]%
        {vossing2022designing}
\bibfield{author}{\bibinfo{person}{Michael V{\"o}ssing},
  \bibinfo{person}{Niklas K{\"u}hl}, \bibinfo{person}{Matteo Lind}, {and}
  \bibinfo{person}{Gerhard Satzger}.} \bibinfo{year}{2022}\natexlab{}.
\newblock \showarticletitle{Designing transparency for effective human-AI
  collaboration}.
\newblock \bibinfo{journal}{\emph{Information Systems Frontiers}}
  \bibinfo{volume}{24}, \bibinfo{number}{3} (\bibinfo{year}{2022}),
  \bibinfo{pages}{877--895}.
\newblock


\bibitem[Wang and Yin(2021)]%
        {wang2021explanations}
\bibfield{author}{\bibinfo{person}{Xinru Wang} {and} \bibinfo{person}{Ming
  Yin}.} \bibinfo{year}{2021}\natexlab{}.
\newblock \showarticletitle{Are explanations helpful? A comparative study of
  the effects of explanations in AI-assisted decision-making}. In
  \bibinfo{booktitle}{\emph{Proceedings of the 26th International Conference on
  Intelligent User Interfaces}}. \bibinfo{pages}{318--328}.
\newblock


\bibitem[Wilder et~al\mbox{.}(2020)]%
        {wilder2020learning}
\bibfield{author}{\bibinfo{person}{Bryan Wilder}, \bibinfo{person}{Eric
  Horvitz}, {and} \bibinfo{person}{Ece Kamar}.}
  \bibinfo{year}{2020}\natexlab{}.
\newblock \showarticletitle{Learning to complement humans}. In
  \bibinfo{booktitle}{\emph{Proceedings of the Twenty-Ninth International Joint
  Conference on Artificial Intelligence}}. \bibinfo{pages}{1526--1533}.
\newblock


\bibitem[Xiong~Chen and Aryee(2007)]%
        {xiong2007delegation}
\bibfield{author}{\bibinfo{person}{Zhen Xiong~Chen} {and}
  \bibinfo{person}{Samuel Aryee}.} \bibinfo{year}{2007}\natexlab{}.
\newblock \showarticletitle{Delegation and employee work outcomes: an
  examination of the cultural context of mediating processes in China}.
\newblock \bibinfo{journal}{\emph{Academy of Management Journal}}
  \bibinfo{volume}{50}, \bibinfo{number}{1} (\bibinfo{year}{2007}),
  \bibinfo{pages}{226--238}.
\newblock


\bibitem[Zhang et~al\mbox{.}(2017)]%
        {zhang2017leaders}
\bibfield{author}{\bibinfo{person}{Xiyang Zhang}, \bibinfo{person}{Jing Qian},
  \bibinfo{person}{Bin Wang}, \bibinfo{person}{Zhuyun Jin},
  \bibinfo{person}{Jiachen Wang}, {and} \bibinfo{person}{Yu Wang}.}
  \bibinfo{year}{2017}\natexlab{}.
\newblock \showarticletitle{Leaders’ behaviors matter: the role of delegation
  in promoting employees’ feedback-seeking behavior}.
\newblock \bibinfo{journal}{\emph{Frontiers in Psychology}}
  \bibinfo{volume}{8} (\bibinfo{year}{2017}), \bibinfo{pages}{1--10}.
\newblock


\bibitem[Zhang et~al\mbox{.}(2020)]%
        {zhang_efect_2020}
\bibfield{author}{\bibinfo{person}{Yunfeng Zhang}, \bibinfo{person}{Q~Vera
  Liao}, {and} \bibinfo{person}{Rachel~KE Bellamy}.}
  \bibinfo{year}{2020}\natexlab{}.
\newblock \showarticletitle{Effect of confidence and explanation on accuracy
  and trust calibration in AI-assisted decision making}. In
  \bibinfo{booktitle}{\emph{Proceedings of the Conference on Fairness,
  Accountability, and Transparency}}. \bibinfo{pages}{295--305}.
\newblock


\end{thebibliography}

\end{document}